\documentclass[prd,preprint,a4paper,showpacs,showkeys]{revtex4}
\usepackage{amsmath,amssymb}
\usepackage{latexsym}
\usepackage{bm}

\newcommand*{\p}{\partial}

\newcommand*{\mm}{{\cal V}}       
\newcommand*{\D}{D}

\newcommand*{\mmI}{\mm^I}
\newcommand*{\mmE}{\mm^E}

\newcommand*{\mmIo}{\mm^I_0}
\newcommand*{\mmIEo}{\mm^{I/E}_0}

\newcommand*{\mmIe}{\mm^I_\ee}
\newcommand*{\mmEe}{\mm^E_\ee}

\newcommand*{\De}{D_\ee}      
\newcommand*{\Do}{{D_0}}
\newcommand*{\K}{K}

\newcommand*{\g}{g}

\newcommand*{\gI}{\g^I}
\newcommand*{\gE}{\g^E}

\newcommand*{\gIo}{\g^I_0}
\newcommand*{\gEo}{\g^E_0}

\newcommand*{\gIe}{\g^I_\ee}
\newcommand*{\gEe}{\g^E_\ee}

\newcommand*{\ovg}{{\hat{\gamma}}}

\newcommand*{\spI}{(\mmI,\, \gI)}
\newcommand*{\spE}{(\mmE,\, \gE)}
\renewcommand*{\sp}{(\mm,\, \g)}

\newcommand*{\spIe}{(\mmIe,\, \gIe)}
\newcommand*{\spEe}{(\mmEe,\, \gEe)}
\newcommand*{\spo}{(\mm_0,\, \g_0)}
\newcommand*{\spIo}{(\mmI_0,\, \gI_0)}
\newcommand*{\spEo}{(\mmE_0,\, \gE_0)}

\renewcommand*{\sup}{\Sigma}
\newcommand*{\suup}{\sigma}

\newcommand*{\supI}{\sup^I}
\newcommand*{\supE}{\sup^E}
\newcommand*{\supIE}{\sup^{I/E}}

\newcommand*{\supe}{\sup_\ee}
\newcommand*{\supIe}{\sup^I_\ee}
\newcommand*{\supEe}{\sup^E_\ee}

\newcommand*{\suupe}{\suup_\ee}
\newcommand*{\supIo}{\sup^I_0}
\newcommand*{\supEo}{\sup^E_0}
\newcommand*{\supIEo}{\sup^{I/E}_0}
\newcommand*{\supo}{\sup_0}
\newcommand*{\suupo}{\suup_0}

\newcommand*{\xI}{\bar{x}_I}

\newcommand*{\U}{U_{\ee}}
\newcommand*{\A}{A_{\ee}}

\newcommand*{\F}{F}
\newcommand*{\Fe}{F_{\ee}}

\renewcommand*{\k}{k_{\ee}}
\newcommand*{\Omegae}{\Omega_\ee}

\newcommand*{\Omegarho}{\Omega_{,\rho}}
\newcommand*{\Omegaz}{\Omega_{,z}}

\renewcommand*{\H}{H}
\newcommand*{\J}{J}
\newcommand*{\G}{V}
\renewcommand*{\L}{\vec{n}V}
\newcommand*{\Y}{W}
\newcommand*{\W}{\vec{n}W}

\newcommand*{\Usupe}{\G_{\ee}}
\newcommand*{\nUsupe}{\L_{\ee}}

\newcommand*{\Omegaesupe}{\Y_{\ee}}
\newcommand*{\nOmegaesupe}{\W_{\ee}}

\newcommand*{\Fsupe}{\H_{\ee}}
\newcommand*{\nFsupe}{\J_{\ee}}

\newcommand*{\Usupo}{\G_{0}}
\newcommand*{\nUsupo}{\L_{0}}

\newcommand*{\Omegasupo}{\Y_{0}}
\newcommand*{\nOmegasupo}{\W_{0}}

\newcommand*{\Fsupo}{\H_{0}}
\newcommand*{\nFsupo}{\J_{0}}

\newcommand*{\Usupprime}{\G'_{0}}
\newcommand*{\nUsupprime}{\L'_{0}}

\newcommand*{\Omegasupprime}{\Y'_{0}}
\newcommand*{\nOmegasupprime}{\W'_{0}}

\newcommand*{\Fsupprime}{\H'_{0}}
\newcommand*{\nFsupprime}{\J'_{0}}

\newcommand*{\Usupprimeprime}{\G''_{0}}
\newcommand*{\nUsupprimeprime}{\L''_{0}}

\newcommand*{\Omegasupprimeprime}{\Y''_{0}}
\newcommand*{\nOmegasupprimeprime}{\W''_{0}}

\newcommand*{\Fsupprimeprime}{\H''_{0}}
\newcommand*{\nFsupprimeprime}{\J''_{0}}

\renewcommand*{\P}{P}
\newcommand*{\Q}{Q}
\newcommand*{\X}{X}

\newcommand*{\QQ}{S}

\newcommand*{\fo}{\F_0 |_{\supo}}
\newcommand*{\efo}{\partial_{\mu} (\F_0 |_{\supo})}
\newcommand*{\nfo}{\vec{n} (\F_0 ) |_{\supo}}
\newcommand*{\eefo}{\partial_{\mu \mu} (\F_0 |_{\supo})}
\newcommand*{\enfo}{\partial_{\mu} (\vec{n} (\F_0 ) |_{\supo})}
\newcommand*{\nnfo}{ \left (n^i n^j \p_i \p_j \F_0 |_{\supo} \right )}

\newcommand*{\eenfo}{\partial_{\mu \mu} (\vec{n} (\F_0 ) |_{\supo})}
\newcommand*{\ennfo}{\partial_{\mu} \left ( n^i n^j \p_i \p_j \F_0 |_{\supo} \right ) }
\newcommand*{\nnnfo}{\left ( n^i n^j n^k \p_i \p_j \p_k \F_0 |_{\supo} \right ) }

\newcommand*{\UEo}{U_0 |_{\supEo}}
\newcommand*{\eUo}{\partial_{\mu} (U_0 |_{\supo})}
\newcommand*{\nUo}{\vec{n} (U_0 ) |_{\supo}}
\newcommand*{\nUEo}{\vec{n} (U_0 ) |_{\supEo}}
\newcommand*{\eeUo}{\partial_{\mu \mu} (U_0 |_{\supo})}
\newcommand*{\enUo}{\partial_{\mu} (\vec{n} (U_0 ) |_{\supo})}
\newcommand*{\nnUo}{ \left (n^i n^j \p_i \p_j U_0 |_{\supo} \right )}

\newcommand*{\nnnUo}{\left ( n^i n^j n^k \p_i \p_j \p_k U_0 |_{\supo} \right ) }

\newcommand*{\Uprimeo}{U'_0 |_{\supo}}
\newcommand*{\UprimeEo}{U'_0 |_{\supEo}}
\newcommand*{\eUprimeo}{\partial_{\mu} (U'_0 |_{\supo})}
\newcommand*{\nUprimeo}{\vec{n} (U'_0 ) |_{\supo}}
\newcommand*{\nUprimeEo}{\vec{n} (U'_0 ) |_{\supEo}}
\newcommand*{\eeUprimeo}{\partial_{\mu \mu} (U'_0 |_{\supo})}
\newcommand*{\nnUprimeo}{ \left (n^i n^j \p_i \p_j U'_0 |_{\supo} \right )}

\newcommand*{\Uprimeprimeo}{U''_0 |_{\supo}}
\newcommand*{\UprimeprimeEo}{U''_0 |_{\supEo}}

\newcommand*{\nUprimeprimeo}{\vec{n} (U''_0 ) |_{\supo}}
\newcommand*{\nUprimeprimeEo}{\vec{n} (U''_0 ) |_{\supEo}}

\newcommand*{\Omegaprimeo}{\Omega'_0 |_{\supo}}

\newcommand*{\OmegaprimeEo}{\Omega'_0 |_{\supEo}}
\newcommand*{\eOmegaprimeo}{\partial_{\mu} (\Omega'_0 |_{\supo})}
\newcommand*{\nOmegaprimeo}{\vec{n} (\Omega'_0 ) |_{\supo}}
\newcommand*{\nOmegaprimeEo}{\vec{n} (\Omega'_0 ) |_{\supEo}}
\newcommand*{\eeOmegaprimeo}{\partial_{\mu \mu} (\Omega'_0 |_{\supo})}
\newcommand*{\nnOmegaprimeo}%
 { \left (n^i n^j \p_i \p_j \Omega'_0 |_{\supo} \right )}

\newcommand*{\fper}{F'_0}
\newcommand*{\efper}{\partial_{\mu} (F'_0)}
\newcommand*{\nfper}{\vec{n} (F'_0 ) |_{\supo}}

\newcommand*{\nnfper}{\left ( n^i n^j \p_i \p_j F'_0 |_{\supo} \right ) }

\newcommand*{\fperper}{F''_0}
\newcommand*{\nfperper}{\vec{n} (F''_0 ) |_{\supo}}

\newcommand*{\embed}{\chi}
\newcommand*{\embede}{\chi_\ee}

\newcommand*{\rmd}{{\text d}}

\newcommand*{\rs}{r_0}
\newcommand*{\xs}{x_0}

\newcommand*{\pw}{y}

\newcommand*{\ee}{\epsilon}
\newcommand*{\al}{\Upsilon_{\pw}}

\newcommand*{\tang}{\vec{e}}

\newcommand*{\sss}{{{\cal S}_0}}

\newcommand*{\minusnUSch}{\frac{1}{\sqrt{\xs^2-1}}}

\newtheorem{lemma}{Lemma}[section]
\newtheorem{proposition}{Proposition}[section]
\newtheorem{theorem}{Theorem}[section]

\begin{document}

\title{Stationary axisymmetric exteriors for perturbations of isolated
  bodies in general relativity, to second order}

\author{Malcolm A.H. MacCallum}
\email{m.a.h.maccallum@qmul.ac.uk}
\affiliation{School of Mathematical Sciences,
  Queen Mary, University of London,
  Mile End Road, London E1 4NS, U.K.}
\author{Marc Mars}
\altaffiliation[Also at: ]{Laboratori de F\'{\i}sica
  Matem\`atica, Societat Catalana de F\'{\i}sica, IEC, Barcelona}
\email{marc@usal.es; raul_vera@ehu.es}
\affiliation{Facultad de Ciencias, Universidad de Salamanca, 
  Plaza de la Merced s/n, 37008 Salamanca, Spain.}
\author{Ra\"ul Vera}
\altaffiliation[Also at: ]{Laboratori de F\'{\i}sica
  Matem\`atica, Societat Catalana de F\'{\i}sica, IEC, Barcelona}
\email{marc@usal.es; raul_vera@ehu.es}
\affiliation{Fisika Teorikoaren Saila, Euskal Herriko Unibertsitatea,
  644 P.K. 48080 Bilbo, Basque Country, Spain}

\date{\today}

\begin{abstract}
  Perturbed stationary axisymmetric isolated bodies, e.g.\ stars,
  represented by a matter-filled interior and an asymptotically flat
  vacuum exterior joined at a surface where the Darmois matching
  conditions are satisfied, are considered. The initial state is
  assumed to be static.  The perturbations of the matching conditions
  are derived and used as boundary conditions for the perturbed Ernst
  equations in the exterior region. The perturbations are calculated
  to second order.  The boundary conditions are overdetermined:
  necessary and sufficient conditions for their compatibility are
  derived. The special case of perturbations of spherical bodies is
  given in detail.
\end{abstract}



\maketitle

\baselineskip=15pt
\section{Introduction}

One would like to have global solutions for rotating objects in
general relativity consisting of a matter-filled interior region and a
vacuum, asymptotically flat, exterior, with the aim of modelling
planets, stars, star clusters, galactic nuclei or galaxies. The
interior and exterior would be matched across a boundary, the surface
of the object, $\Sigma$. Finding such global models is very difficult
even for axially symmetric configurations in equilibrium.  So far
there are no explicit global models known other than those for spherical stars,
which must be non-rotating, and Neugebauer and Meinel's disc of dust
\cite{NeuMei93}, which has no interior (the matter source
has zero thickness and is described by jumps in the metric
derivatives).

There have been several recent studies of the structure of the
underlying equations for the problem. The two regions can be treated
independently subject to the matching at the boundary. For a given
interior, this fixes ``Cauchy data'' giving both the metric and its
derivative at the boundary, i.e.\ gives both Dirichlet and Neumann
boundary conditions for the elliptic equations governing the vacuum
exterior. A compatibility problem therefore arises.  Uniqueness of the
exterior solution has been shown in \cite{MarSen98,Ver03}
(independent of any non-circularity in the interior), and necessary
conditions on the Cauchy data for the existence of the exterior also
exist \cite{Mar06}.  Correspondingly, the interior must
satisfy these conditions in order to describe an isolated rotating
compact object in equilibrium.  Moreover, the conditions were found to
be sufficient for static exteriors \cite{Mar99}, and it has been
conjectured the same holds for the stationary case, but this issue is
still under investigation \cite{Mar06}.

However, applying these conditions in specific situations has proved
very difficult. For example, we had hoped to be able to use them to
give a definitive treatment of the matching problem for the Wahlquist
solution \cite{BraFodMar00,BraFodPer00,SarHoe06} but were unable to
evaluate the required integrals in closed form.  The understanding of
rotating objects has advanced over the years through two major
approaches, numerical relativity and perturbation theory.  The latter
has proved to be very useful and has been widely employed, leading for
example to a well-developed theory of slowly rotating stars.  Probably
the key paper in the theory of slowly rotating stars is that of Hartle
\cite{Har67} (see also \cite{HarTho68,HarSha66}). He discussed the
case of a rigidly rotating non-singular perfect fluid interior with
reflection symmetry, with perturbations dependent only on the slow
rotation imposed on a static background. The uniformity of rotation
was justified by the argument that configurations minimizing the total
mass-energy must rotate uniformly, and hence so must all stable
configurations \cite{Har67,HarSha66}.  He implicitly assumed spherical
symmetry of the background (an assumption later shown to be true for
many cases \cite{BeiSim92}), thus excluding convective motions in the
interior, and the admissibility and C$^2$ character of the coordinates
used, which is greater differentiability than required by the usual
matchings \cite{MarSen93}.

With the aim of studying models for stars (not necessarily in
equilibrium) using perturbation theory, there has been another
direction of research. This has focused on the study of general
perturbations around configurations constructed by the matching of
static spherically symmetric spacetimes.  The first works in this
direction were the classical papers by Gerlach and Sengupta
\cite{GerSen79,GerSen79a}, whose procedure was better justified by
Mart\'in-Garc\'ia and Gundlach \cite{MarGun01} who considered
gauge-independent quantities. The complete formulation of the relevant
perturbed matching conditions to first order, and in general, was
eventually given in independent papers by Battye and Carter
\cite{BatCar95} and Mukohyama \cite{Muk00}. Since these works are
not in principle aimed at \textit{isolated} stars \textit{in
equilibrium}, there are no restrictions on the exterior due to
asymptotic flatness or stationarity, and only first order
perturbations are considered.  The latter is important, since, as was
pointed out in \cite{Har67}, we need to go to second order for
isolated stars in equilibrium in order to obtain the effects of slow
rotation on the shape of the star.  The conditions to second order
have been obtained only recently by one of us \cite{Mar05}.

Yet another method used for the construction of models of slowly
rotating stars has recently \cite{MarMolRui04,CabMarMol06} produced a
particular model to second order in the approximation for a rigidly
rotating and constant density perfect-fluid interior. The method is
based on a two-parameter perturbation scheme, combining the
post-Minkowskian and slow-rotation approximations, both taken up to
the first non-linear level.  The drawback of the model is inherent to
the method, since there is no rotating exact Newtonian limit.

In this paper we re-examine the bases of the perturbed matching
theory.  An obvious question is, therefore, what new things can be
said at this fundamental level?  In our opinion, there are several
issues that need to be clarified. Firstly, have all the hypotheses
involved in the analysis been spelt out in full detail?  Secondly, if
not, what are they, and are they indispensible? This is especially
relevant in the matching of spacetimes, which lies at the very heart
of any approach dealing with finite objects with boundary, such as
models of stars. Thirdly, most, if not all, of the analyses in the
literature deal with a specific matter model in the interior, usually
a rigidly rotating perfect fluid, and the results do depend strongly
on this assumption.  It is natural to ask how one can develop a theory
which is as independent as possible of the interior matter model (or
even fully independent).  Achieving this is clearly of interest, since
the resulting theory could be applied to many different situations,
from differentially rotating perfect fluids to convective fluids,
viscous fluids, or even totally different matter models like elastic
bodies.   The aim of
this paper is to answer all these questions in detail.

One disadvantage of our approach, of course, is that without assuming a
particular matter model we cannot study important questions like uniqueness or
existence of interior solutions for given boundary data. However, even
with specific interiors, the issue of existence of global models for
self-gravitating isolated rotating objects is a very difficult one and
extremely few results are known. The only general theorem in
relativity of this type to date refers to stationary and axially
symmetric rigidly rotating fluids with spatially compact
sections. Existence of such rotating stars was proven by Uwe Heilig
\cite{Hei95} for relativistic configurations close to Newtonian
models and sufficiently small (but finite) angular velocity. In
perturbation theory the situation is not much better, although the
results by Hartle \cite{Har67} indicate a possible path towards a
rigorous existence theorem of rigidly rotating perturbations of a
fluid around a static spherically symmetric configuration.

We believe that this paper is the first to give a comprehensive and
consistent theory of rotating stationary and axisymmetric objects to
first and second order in perturbation theory from first principles.
We start from the Darmois form of the matching conditions, i.e.\ we do
not \textit{a priori} assume admissibility of particular coordinates,
and specialize this form to the stationary axisymmetric case. We do
not assume that the background about which we perturb is spherically
symmetric (unlike previous work), though we shall give the
specialization to this case, nor, as mentioned above, do we restrict
the type of matter in the interior region. The main restrictions are
that: we consider only perturbations about static solutions; we assume
that the exterior is an asymptotically flat vacuum and does not contain an
ergoregion; and we assume that the axis is everywhere regular.

It would in principle be possible to undertake a parallel study of the
more general case where the initial configuration is stationary rather
than static, but we have not explored this at any length. Since
necessary conditions for the existence of a matching of a given
interior to an asymptotically flat exterior can be found in the fully
non-linear regime for this case, they can also be found for the
perturbations.  However, proving sufficiency, even if possible,
appears to be considerably more difficult.

In section \ref{sec:genmatch} we give the general conditions for
matching, specialized to the stationary axisymmetric case and
expressed using Weyl coordinates in the exterior. In section
\ref{sec:scheme} we introduce the perturbation scheme, in particular
introducing some gauge choices. The perturbations in the exterior up
to second order are considered in section \ref{sec:second}: we give
the perturbed Ernst equations together with the perturbed Cauchy data
conditions resulting from the matching.  Necessary and sufficient
conditions on these data for a solution to the exterior equations are
derived in section \ref{Sectcompatibility} (recall that sufficiency
has not yet been shown for the corresponding full equations).
These conditions are first expressed in terms of
the equality of volume integrals over the exterior and surface
integrals involving the Cauchy data and a principal function on the
surface itself, the latter being defined as a solution to a partial
differential equation (PDE). This form is then reduced to integrals on
the boundary surface by first expressing the integrands as divergences
of one-forms, which themselves are ultimately defined as solutions of
ordinary differential equations (ODE).  The final results are thus
found in the form of integrals over the parameter which runs along the
meridians of the boundary surface.  Lastly we specialize the static
background to the spherically symmetric case in section \ref{sphere}:
here some of the functions can be found explicitly, since the exterior
background must be the Schwarzschild solution. We note that we do not
in this paper use the assumptions of Hartle's work \cite{Har67,HarTho68}: we
intend in a later paper to spell out fully the relationship of his
approach and ours.

\section{Matching with Weyl coordinates}
\label{sec:genmatch}

In our problem the spacetime is composed of two regions with boundary,
namely the interior $\spI$ and the exterior $\spE$ (where $I$ and $E$
stand for interior and exterior from now on).  Each region, as well as
its boundary, is assumed to be stationary and axially symmetric, and
to have a regular axis of symmetry (see e.g \cite{MarSen93a}).
We make no specific assumption on the energy-momentum tensor of
$\spI$:  not even the so-called circularity condition
\cite{Pap66} in the interior will be assumed (which for fluids
without energy flux would be equivalent to the absence of convective
motions \cite{Car69,Mar72}). In other words we shall \textit{not}
assume that the orbits of the stationary axisymmetric isometry group
in the interior region are orthogonally transitive (i.e.\ the metric
might not admit coordinates adapted to the isometries in which the
line-element becomes block diagonal). We do not, however, treat
interiors containing non-gravitational fields which propagate to the
exterior (such as electromagnetic fields, for instance). This means
that the exterior region will be taken to be vacuum. Moreover, all
matter configurations will be taken to be isolated, and the spacetimes
are assumed to be asymptotically flat.

Let us denote by $\supI$ and $\supE$ the respective boundaries of
$\spI$ and $\spE$. For us to be able to identify them, these
hypersurfaces must be diffeomorphic to one another. As usual, this is
best described by taking an abstract three-dimensional manifold
$\suup$ and two embeddings $\embed^I$ and $\embed^E$, where
$\embed^{I/E}: \suup \rightarrow \mm^{I/E}$, such that $\supIE
= \embed^{I/E}(\suup)$.  The point-to-point identification of the
boundaries is $\embed^I\circ(\embed^E)^{-1}$ defined on $\supE$.

We assume also that in general the spacetimes admit no further local
symmetries beyond stationarity and axial symmetry (though when
perturbing we shall allow the background spacetime to be spherically
symmetric). Then the axial Killing vector is uniquely fixed by
demanding that its orbits are closed and the Killing coordinate has
periodicity $2 \pi$.  Let us denote by $\vec{\eta}^{\,I}$ and
$\vec{\eta}^{\,E}$ these unique axial Killing vectors in the interior
and the exterior spacetimes, respectively.  As $\spE$ is
asymptotically flat, it admits a unique Killing vector, denoted by
$\vec{\xi}^{\,E}$, which is unit timelike at infinity. In the interior
region there is no equivalent way of fixing the stationary Killing
vector uniquely before the matching is performed.  Two different
situations might be considered, one in which the interior $\spI$ is
unknown and needs to be determined, and the other where it is
explicitly known.  In either case, we can choose $\vec{\xi}^{\,I}$ to
be the unique Killing vector which matches continuously with
$\vec{\xi}^{\,E}$ (recall that the boundaries are also assumed to be
stationary and axially symmetric so that we have a
``symmetry-preserving matching'' \cite{Ver02}), i.e.\ we can
propagate the exterior Killing vector into the interior, but in the
second case this need not agree with the timelike Killing vector used
in writing $\gI$, so we have to introduce two extra parameters in the
interior metric in order to allow the most general matching. This is
well understood and will not be discussed further here; see
\cite{MarSen98}. The isometry group being two-dimensional and
the orbits of $\vec{\eta}^{\,I/E}$ closed, the Killing vectors
$\vec{\xi}^{\,I/E}$ and $\vec{\eta}^{\,I/E}$ commute
(\cite{BiSch84}, see also \cite{Bar01}). Thus, away from the
symmetry axis, there exist local coordinates $\{ T,\Phi, \xI^A \}$
($A,B \cdots = 2,3$) in $\spI$ such that $\vec{\xi}^{\,I}
=\partial_T$ and $\vec{\eta}^{\,I}=\partial_\Phi$.

For the exterior metric, we do not allow ergoregions (as we aim to
model stars rather than black holes) and the vacuum field equations
then imply the local existence of Weyl coordinates $\{t,\phi,\rho,z
\}$ which satisfy the conditions that (i) they are adapted to the
isometries, i.e.\  $\vec{\xi}^{\,E}= \partial_t$ and
$\vec{\eta}^{\,E}= \partial_\phi$, (ii) $\rho=0$ defines the axis
of symmetry and (iii) the metric $\gE$ takes the local form
\begin{eqnarray}
\rmd s^2_E&=&-e^{2U}\left(\rmd t+ A\, \rmd\phi\right)^2
\label{gE} \\
&& +e^{-2U}\left[e^{2k}\left(\rmd\rho^2+\rmd z^2\right)+
 \rho^2 \rmd\phi^2\right],\nonumber
\end{eqnarray}
where $U$, $A$ and $k$ are functions of $\rho$ and $z$ only.  The
scalar $\rho$ is intrinsically defined as the non-negative solution of
$\rho^2 = - (\vec{\xi}^{\,E} , \vec{\xi}^{\,E} )_{\gE}
(\vec{\eta}^{\,E} , \vec{\eta}^{\,E} )_{\gE} + (\vec{\xi}^{\,E} ,
\vec{\eta}^{\,E})_{\gE}^2$ where $( , )_{g}$ denotes the
scalar product with respect to a metric $g$.  $z$ is also
intrinsically defined as the scalar (unique up to a sign and a
constant shift) such that $\rmd z$ is orthogonal to $\rmd\rho$ (with
respect to the metric $\gE$) and $\rmd z$ and $\rmd\rho$ have the same
norm. The coordinate freedom in (\ref{gE}) consists only of constant
shifts of $t$, $\phi$ and $z$. For later use, let us recall that
stationary and axially symmetric vacuum spacetimes admit, locally, a
scalar function $\Omega$ (the twist potential) defined, up to an
additive constant, by
\begin{equation}
\rho \Omegarho= -e^{4 U} A_{,z}, \qquad
 \rho\, \Omegaz =e^{4U}\, A_{\rho}.
\end{equation}
If, moreover, $\spE$ is simply connected $\Omega$ can be defined
globally. In asymptotically flat spacetimes the additive constant is
fixed so that $\Omega \rightarrow 0$ at spatial infinity.  The scalars
$U$ and $\Omega$ are intrinsically defined by $e^{2 U} = - (
\vec{\xi}^{\,E}, \vec{\xi}^{\,E} )$ and $\rmd \Omega=\star(\bm\xi^{\,E}
\wedge\rmd\bm\xi^{\,E})$, where $\bm{\xi^{\,E}} = \gE ( \vec{\xi}^{\,E},
\cdot )$ and $\star$ denotes the Hodge dual \footnote{We shall be using
Hodge duals with respect to several metrics in this paper. The metric
being used in each case should be clear from the context.}.  The
Einstein vacuum field equations reduce to a complex second order PDE
for $(U,\Omega)$ (a pair of real PDEs in the form used below), the
Ernst equation(s), together with quadratures for the metric functions $A$
and $k$.

With this choice of coordinates in $\spE$, and since the stationary and
axial Killing vectors are tangent to the boundary $\supI$, it follows
that there exist local coordinates $\{ \tau,\varphi,\mu \}$ on the
abstract manifold $\suup$ and functions $T (\mu)$, $\Phi(\mu) $ and
$\xI (\mu)$ (see \cite{Ver03}, and \cite{MarSen98} for the
orthogonally transitive case), such that the embedding $\embed^I$
reads
\[
 \embed^I:\{T=\tau + T (\mu) ,\Phi=\varphi + \Phi (\mu) ,
 \xI = \xI (\mu) \}
\]
and, moreover, $\rmd \embed^E ( \partial_\tau ) = \vec{\xi}^{\,I}
|_{\supI}$, $\rmd \embed^E ( \partial_\varphi )
= \vec{\eta}^{\,I} |_{\supI}$ and $\rmd \embed^E ( \partial_\mu )$
is orthogonal to $\vec{\xi}^{\,I}$ and $\vec{\eta}^{\,I}$
on $\supI$.  It can easily be checked that $\{ \tau, \varphi, \mu \}$
is defined uniquely except for constant shifts of $\tau$ and $\varphi$
and redefinitions $\mu (\mu')$.  In terms of these coordinates on
$\suup$, the exterior embedding $\embed^E$ is forced by the matching
conditions (more precisely, by the continuity of the first fundamental
form) to take the following form
\[
 \embed^E:\{t=\tau,\phi=\varphi,\rho=\rho (\mu),z=z (\mu)\}
\]
for some functions $\rho(\mu)$ and $z(\mu)$. Constant shifts in $t$
and $\phi$ are in principle allowed, but they can be set to zero
without loss of generality by exploiting the freedom in the exterior
coordinates. With the embeddings at hand, we can discuss the necessary
and sufficient conditions for the interior and exterior spacetimes to
be joinable (i.e.\ to produce a spacetime $\sp$ with continuous metric
and no surface layers in the Riemann tensor \cite{MarSen93}).
These are the well-known Darmois conditions \cite{Dar27} which
require that the first and second fundamental forms inherited on
$\suup$ from both sides agree. In our case, this set of matching
conditions can be conveniently rewritten as follows (see
\cite{MarSen93b,Mar95}, 
\cite{MarSen98} for the case where the interior is orthogonally
transitive, and \cite{Ver03} for the generalization to an arbitrary
interior):

\begin{enumerate}
\item[(a)] Conditions on the interior hypersurface, given by the Israel
conditions \cite{Isr66}
\begin{eqnarray}
  \label{eq:israel}
  n^{I \alpha} n^{I \beta} S^{I}_{\alpha\beta}~|_{\supI} =0,&&
  n^{I \alpha} {e_{i}}^{I \beta} S^{I}_{\alpha\beta}~|_{\supI}
  =0,\\
  &&i=1,2,3,\nonumber
\end{eqnarray}
where $S^{I}_{\alpha\beta}$ is the Einstein tensor of $\gI$, $\vec
n^I$ is a normal vector to $\supI$ in $\spI$ and $\vec {e}^{\,I}_{i}$
are any three independent vectors tangent to $\supI$. In principle,
any choice of normal vector $\vec{n}^I$ is suitable for writing
(\ref{eq:israel}). However, for the remaining matching conditions it
is convenient to fix this vector so that it points to the interior of
$\spI$ and has norm
\begin{eqnarray}
\left . \left ( \vec{n}^I, \vec{n}^I \right )_{\gI} \right 
 |_{\supI}
 =\left . \left ( \vec{e}^{\,I}, \vec{e}^{\,I} \right )_{\gI} \right 
 |_{\supI},
\label{norm}
\end{eqnarray}
where $\vec{e}^{\,I} = \rmd \embed^I ( \partial_{\mu} )$. Note that
redefining $\mu$ on $\suup$ changes $\vec{e}^{\,I}$ and hence
$\vec{n}^I$. Nevertheless, all expressions below are easily checked to
be invariant under rescalings $\mu ( \mu')$.

Conditions (\ref{eq:israel}) determine which (if any) hypersurfaces in
a given interior metric are candidates to match to an empty exterior.
If the matter content inside is, for instance, a perfect fluid with no
convective motion, these conditions reduce to $p=0$, where $p$ is the
pressure of the fluid. Generically (\ref{eq:israel}) consists of four
independent conditions, which means that only for special interior
metrics will hypersurfaces matching with vacuum exist.  Whenever the
circularity condition is satisfied in the interior, two of
(\ref{eq:israel}) become trivial.

\item[(b)] Definition of the exterior matching hypersurface.
The functions $\rho(\mu)$ and $z (\mu)$ determining the form
of the exterior surface are uniquely fixed by
\begin{equation}
\label{eq:emh}
\rho (\mu)=\alpha |_{\supI},~~~~
 \dot z (\mu)=- \vec n^I (\alpha )|_{\supI},
\end{equation}
where $\alpha^2 \equiv - (\vec{\xi}^{\,I} , \vec{\xi}^{\,I} )_{\gI}
(\vec{\eta}^{\,I} , \vec{\eta}^{\,I} )_{\gI} + (\vec{\xi}^{\,I} ,
\vec{\eta}^{\,I} )_{\gI}^2$, $\alpha \geq 0$, and the dot denotes the
derivative with respect to $\mu$.  The additive constant in $z(\mu)$
is inessential given the shift freedom $z\to z+\mbox{const.}$

\item[(c)] Boundary conditions for the exterior problem. The rest
of the matching conditions provide the following
data on the exterior metric functions $U$ and $A$ on $\supE$
\begin{eqnarray}
U|_{\supE}= V|_{\supI},& \vec n^E (U)|_{\supE}=\vec n^I (V)|_{\supI},
\label{eq:condc1} \\
A|_{\supE}= B|_{\supI},& \vec n^E (A)|_{\supE}=\vec n^I (B)|_{\supI},
\label{eq:condc2}
\end{eqnarray}
where $\vec n^E= -\dot z\,\partial_\rho+\dot
\rho\,\partial_z|_{\supE}$, $e^{2 V} \equiv - (\vec{\xi}^{\,I},
\vec{\xi}^{\,I} )_{\gI}$, and $B = (\vec{\xi}^{\,I}, \vec{\eta}^{\,I}
)_{\gI} / (\vec{\xi}^{\,I}, \vec{\xi}^{\,I} )_{\gI}$.  In order to
ensure that $\vec n^E$ points to the exterior of $\spE$ we choose
$\rho$ to be an increasing function of $\mu$ at the south pole (i.e.\
at the intersection of $\supE$ with the symmetry axis having minimum
value of $z$, which we assume exists). This choice constrains the
allowed rescalings $\mu(\mu')$ to be strictly increasing.

The conditions for $A$ translate into boundary conditions
for the twist potential as follows
\begin{eqnarray}
\dot\Omega|_{\supE}&=&\Omegarho \,\dot\rho+ \Omegaz \,\dot z|_{\supE}
 \nonumber\\
&=&-\left.{\frac{e^{4U}} {\rho}}\vec n^E(A)\right|_{\supE}
 \nonumber\\
&=&-\left.{\frac{e^{4V}} {\alpha }}\vec n^I( B)\right|_{\supI},
 \label{eq:condtwist}\\
\vec n^E(\Omega)|_{\supE}&=&
 -\left.{\frac{e^{4U}}{\rho}}\right|_{\supE}\displaystyle{\frac{d}{d\mu}}
 \left(A|_{\supE}\right)\nonumber \\
&=&-\left.{\frac{e^{4V}}{\alpha}}\right|_{\supI}
 \displaystyle{\frac{d}{d\mu}}\left(B|_{\supI}\right).\nonumber
\end{eqnarray}
The right hand sides of (\ref{eq:condc1}), (\ref{eq:condc2}) and
(\ref{eq:condtwist}) are known once the interior is known. Thus the
matching conditions fix the normal derivative of $\Omega$ on $\supE$
uniquely and $\Omega$ on $\supE$ up to an additive constant. This
constant is in principle relevant as $\Omega$ has been defined so that
it vanishes at infinity. However, it can be proven
\cite{MarSen98} that there is at most one value of the additive
constant for which the exterior vacuum field equations with the
(overdetermined) boundary data (\ref{eq:condc1}) and
(\ref{eq:condtwist}) are compatible.
\end{enumerate}

\section{The perturbed matching conditions}
\label{sec:scheme}

As usual in perturbation theory, we consider a one-parameter family
$(\mm_\ee, g_\ee)$ of 4-dimensional spacetimes, differentiable in
$\ee$, and think of perturbations in terms of derivatives of the
metric with respect to $\epsilon$, evaluated at $\epsilon=0$, which
requires working on a single manifold (more precisely, one manifold
for the interior region and one manifold for the exterior region).
Quantities in the previous section, except for the coordinates, as we
shall discuss below, will now bear a subscript $\ee$.

Let us discuss the interior region (the exterior case will be
similar).  In order to deal with a single manifold, we need to
identify in some way points of different spacetimes $\mmIe$ in the
$\epsilon$-family.  If the manifolds were without boundary and
diffeomorphic to each other we could take any diffeomorphism, smooth
in $\ee$, between, say, $\mmIo$ and $\mmIe$ in order to identify them.
It is clear that such an identification is not unique, and that there
is no canonical choice, because we can perform a diffeomorphism
$\Xi_{\epsilon}$ of $\mmIe$ onto itself, before applying the
diffeomorphism above. This freedom in performing the identification is
the heart of the gauge freedom inherent to perturbation theory (see
e.g.\ \cite{SonBru98}).

Once an identification has been chosen, we have a single manifold
$\mmIo$ and a collection of metrics $\gIe$ defined on it. In
perturbation theory (up to $n$-th order), only the background metric
$\gI_0$ and the first $n$ derivatives of $\gIe$ evaluated at
$\epsilon=0$ are of interest.  Thus we have a background spacetime
$\spIo$ and $n$ symmetric tensor fields, $K^I_a$ ($a=1, \cdots n$),
defined on it (the perturbations).  The freedom in the identification
of spacetimes translates into the gauge freedom in perturbation theory
and is defined by $n$ vector fields on $\mmIo$.  For instance, to
first order, and denoting by $K^I_1$ the first order perturbation
tensor, the gauge freedom is $K^{\prime I}_1 = K^I_1 +
\pounds_{\vec{s}_1} g_0$ where $\vec{s}_1$ is the first order gauge
vector and $\pounds$ denotes the Lie derivative. For higher order
perturbations, the gauge transformations are more complicated as they
also involve all lower order terms.

These issues are all well-understood and would have deserved no
inclusion here except that in our case the manifolds we are
considering are \textit{with boundary}.  How to define the identification
in this case has recently been discussed in \cite{Mar05}. We repeat
the main idea here for completeness. For each $\epsilon$,  $\mmIe$
is a manifold with boundary $\supIe$. Thus  identifying them via 
diffeomorphisms requires, strictly speaking, that boundaries are
mapped into boundaries. However, if we view each manifold $\mmIe$
as a closed subset of a larger manifold without boundary, the condition
that the boundaries are mapped to each other strongly restricts the
gauge freedom (at least near the boundaries) and this may not be suitable
for the problem at hand. It is more convenient to let the boundaries
``move'' freely in the identification. Perturbation tensors can still be
defined everywhere on the background spacetime 
$\mmIo$ with boundary $\supIo$ as follows. For points
away from the background boundary $\supIo$ the usual procedure
obviously works. So we only need to worry about how to define perturbations
on $\supIo$. At each point $p \in \supIo$
and for small enough positive $\epsilon$,
the corresponding identified point in $\mmIe$ will lie on $\supIe$
or move towards the interior of $\mmIe$ or move towards its exterior (within
the larger
manifold). In the first two cases we can define the perturbation tensors
by defining the derivatives as one-sided limits with
$\epsilon \rightarrow 0$, $\epsilon >0$.
In the third case, the point $p$ moves towards
the interior of $\mmIe$ for \textit{negative} values of $\epsilon$. So
derivatives can again be defined if we take the one-sided limits with
$\epsilon < 0$. Note that this construction is independent of the larger
manifold with boundary, which can be dispensed with altogether.
Hence we can, as before, define
$n$ symmetric perturbation tensors $K^I_a$ on the background manifold
with boundary $\spIo$ up to and including the boundary.
Obviously, exactly the
same considerations hold for the exterior region.

Following the discussion above, we can introduce
coordinates $\{T,\Phi,\xI^A \}$
for each spacetime $\spIe$. In principle, the coordinates themselves
should have an $\epsilon$ label. However, we can make this unnecessary
by propagating the coordinates from $\spo$, using the identifications,
so that points with the
same coordinate values $\{ T,\Phi, \xI^A \}$ in the interior regions
of different spacetimes are identified. This choice reduces the gauge
freedom available since we are only left with the changes of
identification describable by
coordinate changes of
the type $\{ T \rightarrow T + T_1 ( \xI^A, \epsilon),\, \Phi
\rightarrow \Phi + \Phi_1 ( \xI^A, \epsilon),\, \xI^A \rightarrow
x_I^{\prime \, A} (\xI,\ee) \}$.  This (partial) gauge fixing is in
fact very useful in the sequel. In this gauge, let us denote by
$K^I_1$ and $K^I_2$ the first and second perturbations of the metric
tensor on the interior background $\spIo$, with boundary $\supIo$.

In the exterior regions, we analogously relate the spacetimes $\spEe$ so
that points with the same Weyl coordinates $\{ t,\phi, \rho,z \}$ in
different spacetimes are identified. Since the Weyl coordinates are
unique except for constant shifts in $t,\phi$ and $z$, the freedom in
performing the identification is reduced to $t \rightarrow t +
\beta_0(\epsilon )$, $\phi \rightarrow \phi + \beta_1 (\epsilon)$, $z
\rightarrow z + \beta_2 (\epsilon)$.  Thus the gauge freedom in the
exterior region is reduced even more strongly than in the interior,
which is also useful.  The exterior background is therefore $\spEo$
with metric
\begin{eqnarray}
g_0  = -e^{2 U_0} \rmd t^2 + e^{-2 U_0} \left [ e^{2 k_0} \left ( \rmd\rho^2
 + \rmd z^2 \right ) + \rho^2 \rmd\phi^2 \right ],
\label{staticWeyl}
\end{eqnarray}
and the perturbation tensors (up to second order) take the following
form
\begin{widetext}
\begin{eqnarray}
K^E_1&=& -2 e^{2 U_0} U'_0 \rmd t^2 - 2 e^{2 U_0} A'_0 \rmd t \rmd\phi
 + 2 e^{- 2 U_0}
 e^{2k_0} \left ( - U'_{0} + k'_0 \right )
 \left (\rmd\rho^2 + \rmd z^2 \right )  
- 2 e^{-2 U_0} U'_0 \rho^2 \rmd\phi^2,
 \nonumber \\
K^E_2 &=& - 2 e^{2 U_0} \left ( U''_0 + 2 {U'_0}^2 \right )
 \rmd t^2 - 2 e^{2 U_0}
 \left ( A''_0 + 4 A'_0 U'_0 \right)
 \rmd t \rmd\phi - 2 \left [ e^{2 U_0 } {A'_0}^2 \right .
 \label{Weylform} \\
&& \left . \phantom{+}+ e^{-2 U_0} \rho^2 \left ( U''_0 -
 2 {U'_0}^2 \right ) \right ] \rmd \phi^2 + 2 e^{-2U_0} e^{2k_0} \left [
 k''_0 -  U''_0 + 2 \left (k'_0 -U'_0 \right )^2 \right ] \left ( \rmd\rho^2
 + \rmd z^2 \right ), \nonumber
\end{eqnarray}
\end{widetext}
which follow directly from (\ref{gE}) by taking $\ee$ derivatives.
Since a gauge choice is involved in the definition, we shall call them
perturbation tensors in \textit{Weyl form} or in \textit{Weyl gauge}.  In
(\ref{staticWeyl}) and (\ref{Weylform}), $U_0$, $k_0$, $U'_0$, $A'_0$,
$k'_0$, $U''_0$, $A''_0$ and $k''_0$ are functions of $\rho$ and $z$
and are defined by $U_0 = \U |_{\ee=0}$, $U'_0 = \partial_{\ee} \U
|_{\ee=0}$, $U''_0 = \partial^2_{\epsilon} \U |_{\ee=0}$, etc.

\section{The second order perturbed exterior problem}
\label{sec:second}

Let us start by recalling the vacuum field equations for (\ref{gE}).
As already mentioned, we only need to concentrate on the equations for
$\U$ and $\Omegae$, since the remaining field equations for $\k$ and
$\A$ reduce to quadratures.

Let us consider Euclidean space $(\mathbb{E}^3, \gamma)$ in
cylindrical coordinates, so $\gamma = \rmd\rho^2 + \rmd z^2 + \rho^2
\rmd\phi^2$. For each value of $\epsilon$, let us consider an axially
symmetric surface $\supe$ defined by $\chi_{\ee} = \{ \rho
=\rho_\ee(\mu), z = z_\ee(\mu), \phi = \varphi \}$ with the same
functions $\rho_\ee(\mu)$ and $z_{\ee}(\mu)$ as those defining
$\supEe$ in $\mmEe$.
Let us denote by
$\De \subset \mathbb{E}^3$  the exterior region of $\supe$.
The use of Weyl coordinates in $\spEe$ allows us to consider
$\{\U,\Omegae \}$ as fields on $\De$. The vacuum field equations
are equivalent to the so-called Ernst equations on $\De$
\begin{equation} \label{eq:ernst}
  \begin{array}[c]{l}
    \displaystyle{\triangle_{\gamma} \U+\frac{1}{2}e^{-4\U}
    \left(\rmd \Omegae,\rmd \Omegae\right)_\gamma=0},\\
    \displaystyle{\triangle_{\gamma}\Omegae-4\left(\rmd \Omegae,\rmd 
      \U\right)_\gamma =0},
  \end{array}
\end{equation}
where $\triangle_{g}$ is the Laplacian of the metric $g$.  This set of
equations is supplemented with the asymptotic values $\U \rightarrow
1$, $\Omegae \rightarrow 0$ at infinity plus the boundary data on
$\supe$ coming from the matching conditions. As already mentioned,
these data
determine both $\{\U,\Omegae \}$ on $\supe$ and its normal derivatives
(except for an additive constant in $\Omegae$).  Thus we are dealing
with Cauchy data for the elliptic system of equations
(\ref{eq:ernst}).  This is an overdetermined problem and we should not
expect solutions to exist for arbitrary data. That expresses the fact
that given an arbitrary stationary and axially symmetric interior
metric (even if it is perfect fluid, say), there will in general be no
stationary and axially symmetric vacuum exterior solution matching
with it \textit{and also asymptotically flat}. Thus existence for the
exterior problem is an important issue. This being true for all
$\epsilon$, the same will happen for the perturbed matching and field
equations, as we discuss next.  In the following subsections, we first
derive the perturbed field equations (subsection \ref{fieldEqs}) and then
the boundary conditions (subsection \ref{Cauchy}).  Section
\ref{Sectcompatibility} is devoted to discussing under what conditions
the Cauchy data of the perturbed field equations are compatible.

\subsection{The Ernst equations up to second order}
\label{fieldEqs}

Let us obtain the systems of equations satisfied by the different
orders in $\ee$ of $\U$ and $\Omegae$.  Differentiating
(\ref{eq:ernst}) with respect to $\ee$ we get
\begin{equation}
  \label{eq:1ernst}
  \begin{array}[c]{l}
    \displaystyle{\triangle_{\gamma}\U'+e^{-4\U}
      \left(\rmd \Omegae,\rmd \Omegae'\right)_\gamma
      -2e^{-4\U}\U' \left(\rmd \Omegae,\rmd \Omegae\right)_\gamma =0},\\[5pt]
    \displaystyle{\triangle_{\gamma}\Omegae'
      -4\left(\rmd \Omegae',\rmd \U\right)_\gamma
      -4\left(\rmd \Omegae,\rmd \U'\right)_\gamma}=0.
  \end{array}
\end{equation}
Differentiating once more we obtain
\begin{widetext}
\begin{eqnarray}
\displaystyle{\triangle_{\gamma}\U''+e^{-4\U}
      \left[-8 \U'\left(\rmd \Omegae,\rmd \Omegae'\right)_\gamma
        +\left(\rmd \Omegae,\rmd \Omegae''\right)_\gamma+
        \left(\rmd \Omegae',\rmd \Omegae'\right)_\gamma\right.}
&&\nonumber \\
\displaystyle{
  \left.-2U''
        \left(\rmd \Omegae,\rmd \Omegae\right)_\gamma
        +8 {U'}^2\left(\rmd \Omegae,\rmd
      \Omegae\right)_\gamma\right]}&=&0,
  \label{eq:2ernst}\\
\displaystyle{\triangle_{\gamma}\Omegae''
      -8\left(\rmd \Omegae',\rmd \U'\right)_\gamma
      -4\left(\rmd \Omegae'',\rmd \U\right)_\gamma
      -4\left(\rmd \Omegae,\rmd \U''\right)_\gamma}&=&0.\nonumber
\end{eqnarray}
\end{widetext}
The systems for the zeroth, first and second order are now obtained
by evaluating the systems (\ref{eq:ernst}), (\ref{eq:1ernst}) and
(\ref{eq:2ernst}) at $\ee=0$.
As already mentioned, we are interested in studying perturbations of static 
objects. Thus
the background exterior metric satisfies $\Omega_0=0$
and hence $U_0$ is a solution of the Laplace equation
\begin{eqnarray}
\triangle_{\gamma}U_0=0. \label{eq:ernst0}
\end{eqnarray}
In fact we are primarily interested in spherical backgrounds so that the 
exterior background metric is
the Schwarzschild metric and $U_0$ is the corresponding Schwarzschild 
solution. Nevertheless, for the sake of
generality we shall keep an arbitrary static and axially symmetric 
background until explicitly stated.
The equations for the first-order perturbation of the
static background $\{U'_0,\,\Omega'_0\}$, which will be called
first-order linearized Ernst equations, read
\begin{equation}
  \label{eq:1ernst0}
  \begin{array}[c]{l}
    \displaystyle{\triangle_{\gamma}U'_0 =0,}\\
    \displaystyle{\triangle_{\gamma}\Omega'_0-
      4\left(\rmd \Omega'_0,\rmd U_0\right)_\gamma=0,}
  \end{array}
\end{equation}
while for the second order perturbation
$\{U''_0,\,\Omega''_0\}$
we get the second order perturbed Ernst equations
\begin{equation}
  \label{eq:2ernst0}
  \begin{array}[c]{l}
    \displaystyle{\triangle_{\gamma}U''_0+e^{-4U_0}
      \left(\rmd \Omega'_0,\rmd \Omega'_0\right)_\gamma=0,}\\
    \displaystyle{\triangle_{\gamma}\Omega''_0
      -4\left(\rmd \Omega''_0,\rmd U_0\right)_\gamma
      -8\left(\rmd \Omega'_0,\rmd U'_0\right)_\gamma=0.}
  \end{array}
\end{equation}

\subsection{Cauchy boundary data up to second order}
\label{Cauchy}

In order to obtain the boundary data for the exterior problem we
should, in a general setting, consider two background spacetimes
$\spIo$ and $\spEo$ which match across the unperturbed boundaries
$\supIo$ and $\supEo$ according to the standard matching conditions.
Moreover, the first and second order perturbed metric tensors
$K_a^{I}$ and $K^E_a$ ($a=1,2$) should satisfy suitable conditions on
$\supIo$ and $\supEo$ coming from suitable first and second
$\epsilon$-derivatives of the full matching conditions.  These
conditions may be called first and second order perturbed matching
conditions.  Their explicit form in full generality has been obtained
by one of us \cite{Mar05}. These can
be specialized to find the perturbed matching conditions 
of interest in this paper. However, for the sake of self-consistency
we shall follow an alternative 
procedure which is well adapted to the stationary and axially
symmetric problem we have at hand. Since, as we saw in Sect.\
\ref{sec:scheme}, the boundary conditions for all $\epsilon$
reorganize themselves into an elegant form and the exterior problem
reduces to the Ernst equation for $(\U,\,\Omegae)$ alone, we need only
concentrate on the perturbed boundary conditions for these objects.
Note that we have introduced coordinates $\{ \tau,\, \varphi,\, \mu \}$,
with no $\epsilon$ label, on the abstract matching manifold $\suupe$.
The reason is the same as before, i.e.\ we identify points in different
$\suupe$ having the same coordinate values $\{ \tau,\, \varphi,\, \mu \}$.

Given the interior family of metrics, the matching conditions give us
(for every $\epsilon$) two functions $\rho_{\ee}(\mu)$ and
$z_{\ee}(\mu)$. Thus we have a family of axially symmetric surfaces
$\supe$ in $\mathbb{E}^3$, all of them diffeomorphic to $\suupo$.
To avoid repetition, we collect together here the assumptions and
notation we shall use.

\textbf{Assumptions:} \textit{Let $\spIo$ and $\spEo$ be two
static and axially symmetric spa\-ce\-ti\-mes which can be joined
across their static and axially symmetric boundaries $\supIo$,
$\supEo$.  Let $\spEo$ be vacuum and asymptotically flat and choose,
locally, Weyl coordinates $(\ref{staticWeyl})$.  Let $K^{I/E}_1$ and
$K^{I/E}_2$ be symmetric tensors on $\mmIEo$, invariant under the
static and axial isometries and such that $K^E_{1/2}$ take the Weyl
form $(\ref{Weylform})$.  Take any metric $\gIe$ in $\mmIo$ such that
$\gIe = \gIo + \epsilon K^I_1 + \frac{1}{2} \epsilon^2 K^I_2 + O
(\epsilon^3)$ and denote by $\vec{\xi}^{\,I}_{\ee}$ and
$\vec{\eta}^{\,I}_{\ee}$ its stationary and axial Killing vectors. Let
$\spIe$ admit a hypersurface $\supIe$ where it can be locally matched
to a vacuum exterior.  Let the identification of $\supIE_\ee$ with
$\suupo$ be made through the common use of coordinates
$\{\mu,\,\varphi\}$ in both manifolds.  On $\suupo$ let functions
$\rho'_0(\mu)$, $\rho''_0(\mu)$, $z'_0(\mu)$, $z''_0(\mu)$ be defined
by $\rho_{\ee}(\mu) = \rho_0 + \epsilon \rho'_0 + \frac{1}{2}
\epsilon^2 \rho''_0 + O (\epsilon^3) =\alpha_{\ee} |_{\supIe}$,
$z_{\ee}(\mu)= \dot{z}_0 + \epsilon \dot{z}'_0 + \frac{1}{2}
\epsilon^2 \dot{z}''_0 + O (\epsilon^3) = - \vec{n}^{I}_{\ee} (
\alpha_{\ee} ) |_{\supIe},~($where $\alpha_{\ee}$ and
$\vec{n}^{\,I}_{\ee}$ are as in $(\ref{norm})$ and $(\ref{eq:emh})$,
evaluated at $\supe)$, and
functions $\Usupe$, $\nUsupe$, $\Omegaesupe$, $\nOmegaesupe$,
$\Usupo(\mu)$, $\Usupprime(\mu)$, $\Usupprimeprime(\mu)$,
$\nUsupo(\mu)$, $\nUsupprime(\mu)$, $\nUsupprimeprime(\mu)$,
$\Omegasupo(\mu)$, $\Omegasupprime(\mu)$, $\Omegasupprimeprime(\mu)$,
$\nOmegasupo(\mu)$, $\nOmegasupprime(\mu)$,
$\nOmegasupprimeprime(\mu)$ by
} 
\begin{widetext}
\begin{eqnarray}
\Usupe=\Usupo + \epsilon \Usupprime + \frac{1}{2} \epsilon^2 \Usupprimeprime +
 O (\epsilon^3) &=& \left . \frac{1}{2} \log \left ( 
 - (\vec{\xi}^{\,I}_\ee, \vec{\xi}^{\,I}_\ee )_{\gIe} \right )  
 \right |_{\supIe} , \nonumber \\
\nUsupe=\nUsupo + \epsilon \nUsupprime + \frac{1}{2} \epsilon^2
 \nUsupprimeprime + 
 O (\epsilon^3) &=& \left . \frac{ \vec{n}^{\,I}_{\ee} 
 (\vec{\xi}^{\,I}_\ee,
 \vec{\xi}^{\,I}_\ee )_{\gIe} }{2 
 (\vec{\xi}^{\,I}_\ee,
 \vec{\xi}^{\,I}_\ee )_{\gIe}} \right |_{\supIe}, \label{epsfns}\\
\dot{\Omegaesupe}= \epsilon \dot{\Omegasupprime} +
 \frac{1}{2} \epsilon^2 
 \Omegasupprimeprime + O (\epsilon^3) &=& \left . 
 - \frac{(\vec{\xi}^{\,I}_\ee, \vec{\xi}^{\,I}_\ee )_{\gIe}^2}{\alpha_{\ee}}
 \vec{n}^{\,I}_{\ee} \left ( \frac{
 (\vec{\xi}^{\,I}_\ee, \vec{\eta}^{\,I}_\ee )_{\gIe}}{
  (\vec{\xi}^{\,I}_\ee, \vec{\xi}^{\,I}_\ee )_{\gIe}} \right )
 \right |_{\supIe}, \nonumber\\
\nOmegaesupe=  \epsilon \nOmegasupprime + \frac{1}{2} \epsilon^2 
 \nOmegasupprimeprime + O (\epsilon^3) &=& \left . 
 - \frac{(\vec{\xi}^{\,I}_\ee, \vec{\xi}^{\,I}_\ee )_{\gIe}^2}{\alpha_{\ee}}
 \frac{ d}{d \mu} 
 \left ( \frac{
 (\vec{\xi}^{\,I}_\ee, \vec{\eta}^{\,I}_\ee )_{\gIe}}{
 (\vec{\xi}^{\,I}_\ee, \vec{\xi}^{\,I}_\ee )_{\gIe}} \right )
 \right |_{\supIe}.\nonumber
\end{eqnarray}
\end{widetext}

Note that the staticity implies that the background values are
$\Omega_0=\Omegasupo(\mu)=0$ and $\vec{n}\Omega_0 =\nOmegasupo(\mu)=0$.
  As shown in the previous section, the matching conditions imply that
$\Usupe$, $\nUsupe$, $\Omegaesupe$ and $\nOmegaesupe$ are provided by
the interior metrics, and give the boundary values $\U |_{\supe}
= \Usupe$, $\vec{n}_{\ee} (\U) |_{\supe} = \nUsupe$, $\Omegae |_{\supe}
= \Omegaesupe$ and $\vec{n}_{\ee} (\Omegae) |_{\supe} = \nOmegaesupe$
for the exterior problem. Note that the right hand sides of the four
equations just given are functions of $\mu$ and $\ee$ alone, while the
left hand sides are functions on $\mathbb{E}^3$ evaluated on a
(moving) surface $\supe$.  We can now take derivatives of these
expressions with respect to $\epsilon$ (at constant $\mu$) in order to
obtain the perturbed boundary conditions for the perturbation
functions $U'_0$, $\Omega'_0$, $U''_0$ and $\Omega''_0$.  It is clear
that when taking derivatives on the right hand sides of (\ref{epsfns})
two types of
terms appear, namely those coming from the explicit dependence on
$\ee$ in the functions and those coming from the fact that the
surfaces $\supe$ also depend on $\ee$.  The latter will involve
$\partial_\rho$ and $\partial_z$ (and higher) derivatives of lower
order terms $U_0$, $\Omega_0$, $U'_0$ and $\Omega'_0$, all of them
evaluated at the unperturbed surface $\sup_0$. Our aim is to express
everything in terms of the twelve functions $\Usupo(\mu)$,
$\Usupprime(\mu)$, $\Usupprimeprime(\mu)$, $\nUsupo(\mu)$,
$\nUsupprime(\mu)$, $\nUsupprimeprime(\mu)$, $\Omegasupo(\mu)$,
$\Omegasupprime(\mu)$, $\Omegasupprimeprime(\mu)$, $\nOmegasupo (\mu)$,
$\nOmegasupprime(\mu)$, and $\nOmegasupprimeprime(\mu)$ of $\mu$,
which correspond exactly to the boundary information coming from the
interior through the background matching and the first and second
order perturbed matching conditions.

Let us take a fixed point $p \in \suupo$ and consider the trajectory
in $\mathbb{E}^3$ defined by $\embede(p)$ when $\epsilon$ varies.  The
first and second derivatives along this trajectory, evaluated at
$\epsilon=0$, give us two vectors on $p$ and hence two vector fields on
$\supo$. They determine how the matching surface moves within
$\mathbb{E}^3$ to second order in approximation theory.  Denoting
these vector fields by $\vec{Z}_1$ and $\vec{Z}_2$, we obviously have
\begin{eqnarray}
\vec{Z}_1 = \rho'_0 \partial_\rho + z'_0 \partial_z |_{\supo}, \qquad
\vec{Z}_2 = \rho''_0 \partial_\rho + z''_0 \partial_z |_{\supo} ,
\label{zdef}
\end{eqnarray}
where, as before, $\rho'_0 \equiv \partial_\ee\rho_\ee|_{\ee=0}$,
$\rho''_0 \equiv \partial_\ee \partial_{\ee} \rho_\ee|_{\ee=0}$, etc.
Two further relevant vectors on $\supo$ are the first and second
perturbations of the normals $\vec{n}_{\ee}$. Explicitly
\begin{eqnarray}
\left . \frac{\partial \vec{n}_{\ee}}{\partial \ee} \right |_{\ee=0}
&=& - \dot{z}'_0 \partial_\rho + \dot{\rho}'_0 \partial_z
|_{\supo},\nonumber\\[-5pt]
\label{dndef} \\[-5pt]
\left . \frac{\partial^2 \vec{n}_{\ee}}{\partial \ee^2} \right |_{\ee=0}
&=& -\dot{z}''_0 \partial_\rho + \dot{\rho}''_0 \partial_z |_{\supo}.\nonumber
\end{eqnarray}
Since we want to rewrite everything in terms of intrinsic objects on
the unperturbed surface $\supo$, we need to use a basis adapted to
this surface. A convenient choice consists of the unperturbed tangent and
normal vectors, namely $\tang \equiv \rmd \embed_0 (\partial_\mu) = \dot
\rho_0 \partial_\rho + \dot z_0 \partial_z |_{\supo}$ and $\vec{n}
=\vec{n}_{\ee} |_{\ee=0}$ (for notational convenience we drop a
subindex 0 both in $\tang$ and in $\vec{n}$). Writing down the vectors
$\{\partial_\rho,\, \partial_z \}$ on $\supo$ in terms of $\tang$ and
$\vec{n}$ we get
\begin{eqnarray}
\partial_z |_{\supo}&=&\frac{1}{\dot z^2_0+\dot\rho^2_0}
\left( \frac{}{} \dot z_0\, \tang +
  \dot \rho_0\,\vec n \right), \nonumber \\[-5pt]
\label{change-basis}\\[-5pt]
  \partial_\rho |_{\supo}&=&\frac{1}{\dot z^2_0+\dot\rho^2_0}
\left( \frac{}{} \dot \rho_0\, \tang -
  \dot z_0\,\vec n \right).\nonumber
\end{eqnarray}
We can now express $\vec{Z}_1$, $\vec{Z}_2$, $\left . \frac{\partial
    \vec{n}_{\ee}}{\partial \ee} \right |_{\ee=0}$, $\left .
  \frac{\partial^2 \vec{n}_{\ee}}{\partial \ee^2} \right |_{\ee=0}$,
in terms of $\{ \tang,\, \vec{n} \}$. When doing this, six scalar fields
on $\supo$ appear in a natural way. Denoting them by $\P_1$, $\Q_1$,
$\P_2$, $\Q_2$, $\X_0$ and $\X_1$ we have
\begin{widetext}
\begin{eqnarray*}
\vec{Z}_1 &=& \P_1 \tang + \Q_1 \vec{n}, \qquad \vec{Z}_2 = \P_2 \tang + \Q_2 
\vec{n}, \\[3pt]
\left . \frac{\partial \vec{n}_{\ee}}{\partial \ee} \right |_{\ee=0} &=&
\left ( -\frac{d \Q_1}{d \mu}  + \P_1 \X_0 - \Q_1 \X_1 \right ) \tang +
\left ( \frac{d \P_1}{d \mu } + \P_1 \X_1 + \Q_1 \X_0 \right ) \vec{n}, \\
\left . \frac{\partial^2 \vec{n}_{\ee}}{\partial \ee^2} \right |_{\ee=0} &=&
\left ( -\frac{d \Q_2}{d \mu}  + \P_2 \X_0 - \Q_2 \X_1 \right ) \tang +
\left ( \frac{d \P_2}{d \mu } + \P_2 \X_1 + \Q_2 \X_0 \right ) \vec{n},
\end{eqnarray*}
where 
\begin{eqnarray}
\P_1 = \frac{\dot{\rho}_0 \rho'_0 + \dot{z}_0 z'_0}{\dot{\rho}_0^2 + 
\dot{z}_0^2 }, \qquad & &
\Q_1 =  \frac{\dot{\rho}_0 z'_0 - \dot{z}_0 \rho'_0}{\dot{\rho}_0^2 + 
\dot{z}_0^2 }, \qquad
\P_2 = \frac{\dot{\rho}_0 \rho''_0 + \dot{z}_0 z''_0}{\dot{\rho}_0^2 + 
\dot{z}_0^2 }, \qquad
\Q_2 = \frac{\dot{\rho}_0 z''_0 - \dot{z}_0 \rho''_0}{\dot{\rho}_0^2 + 
\dot{z}_0^2 }, \nonumber \\ & &
\X_0  =  \frac{\ddot{\rho}_0 \dot{z}_0 - \ddot{z}_0 
\dot{\rho}_0}{\dot{\rho}_0^2 + \dot{z}_0^2 }, \qquad
\X_1 = \frac{\ddot{\rho}_0 \dot{\rho}_0 + \ddot{z}_0 
\dot{z}_0}{\dot{\rho}_0^2 + \dot{z}_0^2 }. \label{PsQs}
\end{eqnarray}
\end{widetext}
Having defined these objects we can evaluate the first and second
order perturbations of the matching conditions. The statement of the
result is rather lengthy.
\begin{widetext}
\begin{proposition}
\label{FullBoundary} Under the Assumptions stated above, the metrics
$\gIe$ and $\gEe \equiv \gEo + \epsilon K^E_1
+\frac{1}{2}\epsilon^2 K^E_2$ match perturbatively to second order on
$\supIEo$ if and only if the following conditions are satisfied
\begin{eqnarray*}
\UEo & = &  \Usupo, \qquad \nUEo = \nUsupo, \qquad
  \UprimeEo = \Usupprime - \P_1 \frac{d \Usupo}{d \mu } - \Q_1 \nUsupo, \\
\nUprimeEo & = &  \nUsupprime + \frac{d}{d \mu} \left ( \Q_1 \frac{d 
  \Usupo}{d \mu} \right ) -
  \frac{d \left ( \P_1 (\nUsupo) \right )  }{d \mu}
  + \Q_1 \left ( \frac{\dot{\rho}_0}{\rho_0} \frac{d \Usupo}{d \mu } - 
  \frac{\dot{z}_0}{\rho_0} \nUsupo \right ) , \\
\UprimeprimeEo & = &  \Usupprimeprime - 2 \P_1 \frac{d \Usupprime}{d \mu } - 
  2 \Q_1  \nUsupprime+
  \frac{d}{d \mu} \left ( \left ( \P_1^2- \Q_1^2 \right ) \frac{d \Usupo}{d 
  \mu} \right )
  + \frac{d \left ( 2 \P_1 \Q_1 \nUsupo \right )}{d \mu}\\
& &
 +\left ( - \P_2 + \P_1^2 \X_1 + 2 \P_1 \Q_1 \X_0 - \Q_1^2 \X_1 - \Q_1^2 
  \frac{\dot{\rho}_0}{\rho_0}
  \right ) \frac{d \Usupo}{d \mu } \\
& & +\left ( - \Q_2 - \P_1^2 \X_0 + 2 \P_1 \Q_1 \X_1 + \Q_1^2 \X_0 + \Q_1^2 
  \frac{\dot{z}_0}{\rho_0}
  \right ) \nUsupo,  \\
\nUprimeprimeEo & = &  \nUsupprimeprime + 2 \frac{d}{d \mu}
  \left( \Q_1  \frac{d \Usupprime}{d \mu} \right)
  - 2 \frac{d \left( \P_1 \nUsupprime \right) }{d \mu}
  + 2 \Q_1 \left( \frac{\dot{\rho}_0}{\rho_0}
  \frac{d \Usupprime}{d \mu }
  - \frac{\dot{z}_0}{\rho_0} \nUsupprime \right)\\
& & - \frac{d^2}{d \mu^2 } \left ( 2 \P_1 \Q_1 \frac{d\Usupo}{d \mu}
  \right )  + \frac{d^2}{d \mu^2} \left ( \frac{}{} \left ( \P_1^2 -
  \Q_1^2 \right ) \nUsupo \right )  \\
& & + \frac{d}{d \mu} \left \{ \left [  \Q_2 
  + \left ( \P_1^2 - \Q_1^2 \right )
  \X_0 -2\P_1\Q_1\frac{\dot{\rho}_0}{\rho_0}  - 2 \P_1 \Q_1 \X_1
  -\Q_1^2 \frac{\dot{z}_0}{\rho_0} \right ]
  \frac{d \Usupo}{d \mu} \right  \}  \\
& &
  + \frac{d}{d \mu} \left \{ \frac{}{}\left [ - \P_2 
  + \left ( \P_1^2  -  \Q_1^2 \right ) \X_1 + 2\P_1 \Q_1
  \frac{\dot{z}_0}{\rho_0} + 2 \P_1 \Q_1 \X_0
   - \Q_1^2 \frac{\dot{\rho}_0}{\rho_0} \right
  ]\nUsupo \right \}  \\
& &
  + \left ( \Q_2 + (\P_1^2 -\Q_1^2)\X_0 - 2 \P_1 \Q_1
  \frac{\dot{\rho}_0}{\rho_0} \right ) \left (
  \frac{\dot{\rho}_0}{\rho_0}  \frac{d \Usupo}{d \mu } - 
  \frac{\dot{z}_0}{\rho_0}  \nUsupo \right ) \\
& &
  + \left ( 2\P_1\Q_1\X_0  - \Q_1^2  \frac{\dot{\rho}_0}{\rho_0}
  \right ) \left (  \frac{\dot{\rho}_0}{\rho_0} \nUsupo
  + \frac{\dot{z}_0}{\rho_0} \frac{d \Usupo}{d \mu } \right ),\\
\OmegaprimeEo &=& \Omegasupprime, \qquad \nOmegaprimeEo= \nOmegasupprime, 
  \qquad
  \Omega''_0 |_{\supEo} = \Omegasupprimeprime-2 \P_1 \frac{d \Omegasupprime}{d 
  \mu}- 2 \Q_1 \nOmegasupprime, \\
\vec{n} \left ( \Omega''_0 \right )  |_{\supEo} & = & \nOmegasupprimeprime +
  2 \frac{d}{d \mu} \left ( \Q_1 \frac{d\Omegasupprime}{d \mu} \right )
  -2 \frac{d\left ( P_1 \nOmegasupprime \right ) }{d \mu}
  \\
& & +2 \Q_1 \left [ \left ( \frac{\dot{\rho}_0}{\rho_0} - 4 \frac{d 
  \Usupo}{d \mu} \right )
  \frac{d \Omegasupprime}{d \mu}  - \left ( \frac{\dot{z}_0}{\rho_0} + 4 
  \nUsupo \right ) \nOmegasupprime \right ].
\end{eqnarray*}
where $\P_1, \P_2, \Q_1, \Q_2, \X_1, \X_2$ are defined in (\ref{PsQs}).
\end{proposition}
\end{widetext}

The actual necessity and sufficiency arises from the corresponding
properties of the general Darmois conditions. The main work in the
proof is the direct but cumbersome calculation needed to
arrive at the formulae: we leave this to the Appendix.

\section{Compatibility conditions}
\label{Sectcompatibility}

In the previous section we found the overdetermined boundary data for
$U'_0$, $\Omega'_0$, $U''_0$, $\Omega''_0$ in terms of the interior
metric and perturbation tensors. Since the equations they satisfy are
elliptic we need to determine under what conditions asymptotically
flat solutions exist. 
Asymptotic flatness requires that, for any value
of $\ee$, $\lim_{\rho^2 + z^2 \rightarrow \infty} \U = 1$ and
$\lim_{\rho^2 + z^2 \rightarrow \infty} \Omegae = 0$.  This implies $
\lim_{\rho^2 + z^2 \rightarrow \infty} U'_0 = \lim_{\rho^2 + z^2
  \rightarrow \infty} \Omega'_0 = \lim_{\rho^2 + z^2 \rightarrow
  \infty} U''_0 = \lim_{\rho^2 + z^2 \rightarrow \infty} \Omega''_0
=0$. 

The problem of determining the Neumann data (i.e.\ the normal derivative
of the function on the boundary $\partial \Omega$ of a domain $\Omega$)
in terms of the Dirichlet data (the value of the function on
$\partial \Omega$), so that the Cauchy problem
for an elliptic equation on $\Omega$ is solvable, is called 
obtaining the Dirichlet-Neumann map. Many results are known on this problem,
including deriving general properties of the map, determining the
coefficients of the elliptic equation from the Dirichlet-Neumann map
and finding explicit representations of this map for especially simple
equations and domains. Good references on this topic are \cite{Fol95}
and \cite{McL00}. Our problem  can then be phrased as saying that we want
to find \textit{explicit} restrictions on the boundary data so that they
solve the Dirichlet-Neumann problem for the linearized Ernst
equations on axially symmetric domains.

While the equation for $U'_0$ is just the Laplace equation in
Euclidean space, the equation for $\Omega'_0$ has, in addition, lower
order terms.  In order to treat all cases at the same time it turns
out to be convenient to define a conformally flat metric $\tilde\gamma
=e^{- 8 U_0} \gamma$ on $\Do$, in terms of which the second equations
in \eqref{eq:1ernst0}-\eqref{eq:2ernst0} can be rewritten as
\begin{eqnarray*}
\triangle_{\tilde\gamma}  \Omega'_0  & =& 0, \\
\triangle_{\tilde\gamma} \Omega''_0  & = & 8
\left(\rmd \Omega'_0,\rmd U'_0\right)_{\tilde\gamma}.
\end{eqnarray*}
Thus all equations for $U'_0$, $U''_0$, $\Omega'_0$, $\Omega''_0$ can be
collectively  written as
\begin{eqnarray}
\triangle_\ovg u = j, \label{master}
\end{eqnarray}
where $u=u(\rho,z)$ stands for $U_0$, $U_0'$, etc..., and $j=j(\rho,z)$
represents the inhomogeneous terms in the second order perturbation
equations. The metric $\ovg$ corresponds to either $\gamma$, for the
$U$-equations, or $\tilde{\gamma}$, for the $\Omega$-equations. The
domain $(\Do,\,\tilde{\gamma})$ is clearly non-compact because
$\tilde{\gamma}$ is an asymptotically flat metric.  Thus the
compatibility conditions for the boundary values of $U'_0$, $U''_0$,
$\Omega'_0$, $\Omega''_0$ can be studied as particular cases of the
compatibility conditions of the Cauchy problem for the general
inhomogeneous Poisson equation (\ref{master}) defined on a non-compact
asymptotically flat region $(\Do,\,\ovg)$ with a boundary $\partial \Do
=\{ \rho = \rho_0 (\mu),\, z = z_0 (\mu),\, \phi = \varphi \}$.
Furthermore, we shall assume that $j$ tends to zero at infinity at
least like $1/r^4$ where $r \equiv \sqrt{ \rho^2 + z^2 }$ (this
requirement is fulfilled in the cases we are concerned with due to
asymptotic flatness).

We start with some well-known facts from potential theory.
A simple consequence of
Gauss' theorem is  Green's identity which, for any
compact domain $\K \subset \Do$ with $C^1$ boundary $\partial \K$ and any
$C^2$ function $\psi$ on $\K$ with $C^1$ extension to $\K \cup
\partial \K$, reads
\begin{eqnarray*}
\int_{\K}\left(\psi \triangle_{\ovg} u- u\triangle_{\ovg} \psi\right)
\eta_{\ovg}
=\int_{\partial \K}\left [ \frac{}{} \psi \vec{n}_{\ovg} ( u ) - u
  \vec{n}_{\ovg} ( \psi ) \right] \rmd S_{\ovg},
\label{Green}
\end{eqnarray*}
where $\vec{n}_{\ovg}$ is a unit (with respect to $\ovg$) normal
vector pointing out from $\K$, $\eta_{\ovg}$ is the volume form of
$(\Do,\,\ovg)$ and $\rmd S_{\ovg}$ is the induced surface element of
$\partial K$. We intend to apply this identity to a function $\psi$
that (i) solves the Laplace equation $\triangle_{\ovg} \psi = 0$ on
$\Do$, (ii) admits a $C^1$ extension to $\partial \Do$ and (iii)
decays at infinity in such a way that $\psi \sqrt{\rho^2 + z^2} $ is a
bounded function on $\Do$. A function $\psi$ satisfying these three
properties is called a \textit{regular $\ovg$-harmonic function} on $\Do$
(if a function satisfies just (ii) and (iii) and is $C^2$ on $\Do$ we
shall call it \textit{regular}). For such a function we can take
$\K = \Do$ in (\ref{Green}) because the integral over the boundary ``at
infinity'' can be easily shown to vanish. Thus
\begin{eqnarray}
\int_{\partial \Do}\left [ \frac{}{} \psi \vec{n}_{\ovg} ( u ) - u
  \vec{n}_{\ovg} ( \psi ) \right] \rmd S_{\ovg} = \int_{\Do} \psi j
  \eta_{\ovg}.
\label{GreenD}
\end{eqnarray}
This expression relates the overdetermined boundary data on $\partial
\Do$ to a volume integral of the inhomogeneous term $j$. Denoting
the boundary data for $u$ by $u |_{\partial Do} \equiv f_0$ and
$\vec{n}_{\ovg} (u) |_{\partial Do} \equiv f_1$, we have,
explicitly,
\begin{eqnarray}
\int_{\partial \Do}\left [ \frac{}{} \psi f_1  - f_0  \vec{n}_{\ovg} (
  \psi ) \right] \rmd S_{\ovg} =  \int_{\Do} \psi j \eta_{\ovg},
\label{GreenBoun}
\end{eqnarray}
which are clearly necessary conditions on the boundary data for
existence of a regular solution $u$.  It is natural to ask whether
such conditions are also sufficient. More precisely, assume that a
continuous function $j$ is given on $\Do$ such that $r^4 j$ is bounded
at infinity. Give also two arbitrary continuous functions $f_0$ and
$f_1$ on $\partial \Do$ which satisfy (\ref{GreenBoun}) for \textit{any}
choice of regular $\ovg$-harmonic function $\psi$. We want to check
whether there always exists a function $u$ satisfying the Poisson
equation $\triangle_{\ovg} u = j$, with $r u$ bounded at infinity and
such that the boundary equations $u |_{\partial \Do} = f_0$,
$\vec{n}_{\ovg} (u) |_{\partial \Do} = f_1$ are satisfied. The answer
is yes as we prove next. Consider the Dirichlet problem
$\triangle_{\ovg} u = j $ with $u |_{\partial \Do} = f_0$. Standard
elliptic theory tells us that this problem always admits a unique
solution $u$ which tends to zero at infinity. Let us define
$\tilde{f}_1$ on $\partial \Do$ by $\tilde{f}_1 \equiv \vec{n}_{\ovg}
(u) |_{\partial \Do}$.  Since $u$ solves the Poisson equation, it
follows from (\ref{GreenD}) that, for any regular $\ovg$-harmonic
function $\psi$,
\begin{eqnarray}
\int_{\partial \Do}\left [ \frac{}{} \psi \tilde{f}_1  - f_0
  \vec{n}_{\ovg} ( \psi ) \right] \rmd S_{\ovg} = \int_{\Do} \psi j \eta_{\ovg}.
\label{GreenBoun2}
\end{eqnarray}
Our assumption is that $f_0$ and $f_1$ satisfy (\ref{GreenBoun}) for
any such $\psi$.  Subtracting (\ref{GreenBoun}) and (\ref{GreenBoun2})
we get $\int_{\partial \Do} \psi ( f_1 - \tilde{f}_1 ) \rmd S_{\ovg}
=0$. However, since we can take any regular $\ovg$-harmonic function
$\psi$, the fact that the Dirichlet problem for the Laplace equation
always admits a solution allows us to choose $\psi |_{\partial \Do}$
to be any continuous function. This readily implies $f_1 = \tilde{f}_1$
and hence that the overdetermined boundary data admits a
decaying solution, as claimed.

However, the compatibility condition (\ref{GreenBoun}) has a
disadvantage, namely that it must be checked for an \textit{arbitrary}
decaying solution $\psi$ of the Laplace equation. This makes it
useless in practical terms. Our aim is to reduce the number of
solutions $\psi$ that must be checked in (\ref{GreenBoun}) in order to
ensure compatibility of $f_0$ and $f_1$.  Here is where axial symmetry
plays an essential role. We restrict ourselves to axially symmetric
functions $f_0$ and $f_1$ on an axially symmetric boundary $\partial
\Do$ and we assume further that $\partial \Do$ is simply connected, so
that it is diffeomorphic to a 2-sphere. Using coordinates $\mu$ and
$\varphi$ on $\partial \Do$, let us denote by $\mu_S$d and $\mu_N$ (with
$\mu_S < \mu_N$) the only two values of $\mu$ at which $\partial \Do$
intersects the axis of symmetry. Let us also assume that they satisfy
$z_0 (\mu_S) < z_0 (\mu_N)$, i.e.\ that $\mu$ increases from the
``south'' pole of the object (corresponding to $\mu_S$) to the
``north'' pole (corresponding to $\mu_N$). With this assumption, a
direct calculation shows that $\rmd S_{\gamma}\vec{n}_{\gamma} = \rho_0
\rmd\mu \rmd\varphi \vec{n} $, where $\vec{n}$ is our usual normal vector
$\vec{n} = - \dot{z}_0 \partial_{\rho} + \dot{\rho}_0 \partial_z
|_{\Do}$.  Similarly $\rmd S_{\tilde{\gamma}} \vec{n}_{\tilde{\gamma}}
= e^{-4 U_0|_{\supo}} \rho_0 \rmd\mu \rmd\varphi \vec{n}$.  For $\ovg
= \gamma$, (\ref{GreenD}) becomes, after a trivial angular integration,
\begin{equation}
  \label{eq:green1}
  \int_{\mu_S}^{\mu_N} \left . \left[\psi\,\vec n (u)-u\,\vec n(\psi)
 \right] \rho_0 \right |_{\supo}  \rmd \mu
= \frac{1}{2 \pi} \int_{\Do}\psi j \eta_{\gamma}.
\end{equation}
For $\ovg = \tilde{\gamma}$,
(\ref{GreenD}) becomes
\begin{eqnarray}
  \int_{\mu_S}^{\mu_N} \left[\psi\,\vec n (u)\right .
&-& \left . \left . u\,\vec n(\psi)
  \right]\rho_0  e^{-4 U_0} \right |_{\supo} \rmd \mu  \nonumber\\
  \label{eq:green2}
&=& \frac{1}{2 \pi}
  \int_{\Do}\psi j e^{-12 U_0 } 
\eta_{\gamma}.
\end{eqnarray}
We want to choose a reduced set of functions $\psi$ for which the
argument used above to show consistency of the data still holds. The following
Lemma is probably known
although we could not find an explicit reference for it.
\begin{lemma}
\label{integral}
Let $h: \left [ \mu_S, \mu_N \right ] \rightarrow \mathbb{R}$ be a
continuous function satisfying
\begin{eqnarray*}
\int_{\mu_S}^{\mu_N} \frac{h(\mu)  \rmd \mu}{\sqrt{\rho_0^2(\mu) +
    (z_0(\mu) - \pw )^2}} =0 
\end{eqnarray*}
for any constant $\pw \in ( z_S,z_N )$, where $z_S \equiv z_0 (\mu_S)$ and 
$z_S \equiv z_0 (\mu_N)$. Then $h \equiv  0$.
\end{lemma}
\textit{Proof:} 
Let us define a function $\tilde{h}$ on $\partial \Do$ by
extending $h$ in an axially symmetric way, i.e.\ $\tilde{h}
(\mu,\varphi) \equiv h (\mu)$. For any point $q \in \mathbb{E}^3$ let
us define the function
\begin{eqnarray*}
Y_{h} (q) = \int_{\partial \Do} \frac{\tilde{h}}{\mbox{dist}(\cdot, q)} 
\rmd S_{\gamma}, 
\end{eqnarray*}
where $\mbox{dist} (\cdot, q)$ denotes Euclidean distance between a
point on $\partial \Do$ and the point $q$. This function, also
called ``single layer potential'', is
well-defined throughout $\mathbb{E}^3$ (including the axis of symmetry
and the surface $\partial \Do$ \cite{Kel53}), 
is axially symmetric and vanishes
at infinity. Moreover, potential theory tells us that $Y_{h}$
is $C^0$ everywhere (including $\partial \Do$) and satisfies
$\triangle_{\gamma} Y_h = 0$ except on $\partial \Do$. The function
$\tilde{h}$ is directly related to the jumps of the first normal
derivative of $Y_h$ on $\partial \Do$ \cite{Kel53}
($Y_h$ can be physically interpreted as
the potential created by a surface layer sitting on $\partial \Do$).
Thus $h$ vanishes if and only if $Y_h$ is $C^1$ on $\mathbb{E}^3$.

The hypothesis of the Lemma tells us that $Y_h$ vanishes on the piece
of the axis lying between the south and the north pole of $\partial
\Do$. Regularity of the Laplace equation shows that $Y_h$ is analytic
except at $\partial \Do$. If $Y_h$ is identically zero in some
neighbourhood of the axis inside $\partial \Do$, then continuity at
$\partial \Do$ and analyticity implies $Y_h \equiv 0$
everywhere and $h=0$ would follow. Let us thus assume that there is a
neighbourhood ${\cal U}$ of a point lying on the axis between the
south and north poles in which $Y_h$ is not identically zero.
Analyticity in Cartesian coordinates implies that $Y_h$ depends
analytically on $ \rho^2$ and $z$ in cylindrical coordinates.  Not
being identically zero, there must exist a minimum value $k \in
\mathbb{N}$ and an analytic (not identically zero) function $g(z)$
such that $Y_h |_{\cal U} = g(z) \rho^{2k} + O(\rho^{2k+2})$.  The
fact that $Y_h$ vanishes on the axis demands $k \geq 1$. Substitution
into the Laplace equation $\partial_{\rho\rho} Y_h + \rho^{-1}
\partial_{\rho} Y_h + \partial_{zz} Y_h = 0$ yields $ 4 k^2 g(z)
\rho^{2k -2} + O(\rho^{2k} )=0$ which is a contradiction. This completes
the proof.$\Box$

This Lemma already suggests which subclass of regular $\ovg$-harmonic
functions needs to considered for the compatibility of the
boundary conditions.  For the flat metric $\gamma$, a natural class of
axially symmetric harmonic functions is
\begin{eqnarray}
\psi_y(\rho,z) \equiv \frac{1}{\sqrt{\rho^2 + (z -y )^2}},
\qquad \pw \in ( z_S,z_N ).
\label{principalU}
\end{eqnarray}
This expression is singular only at $\rho=0, z=y$, which lies on the
axis of symmetry outside $\Do$.  Thus $\psi_y$ is indeed a regular
$\gamma$-harmonic function on $\Do$.

For the $\tilde{\gamma}$ metric (corresponding to the
$\Omega$-equations), we need to find a suitable class of axially
symmetric solutions of $\triangle_{\tilde{\gamma}} u =0$. In order to
find them more easily let us consider the semiplane $\mathbb{K}
= \mathbb{R}^+ \times \mathbb{R}$ defined as the subset $\{ \phi
= \mbox{const.}, \rho \geq 0 \}$ of $\mathbb{E}^3$ in cylindrical
coordinates. Working directly on $\mathbb{K}$ has the advantage that
axial symmetry is incorporated into the calculations from the very
beginning. $\mathbb{K}$ is endowed with a flat metric $\rmd \rho^2 + \rmd
z^2$. We choose the orientation so that $\star \rmd\rho = - \rmd z$ (and
hence $\star \rmd z = \rmd \rho$). Obviously any axially symmetric
function in $\mathbb{E}^3$ immediately defines a function on
$\mathbb{K}$. Similarly a function on $\mathbb{K}$ defines
an axially symmetric function on $\mathbb{E}^3$.
We shall use the same symbol to denote both functions
(the precise meaning should be clear from the context). The field
equations (\ref{eq:1ernst0})-(\ref{eq:2ernst0}) can be translated into
equations on $\mathbb{K}$. It is straightforward to check that the
$\gamma$-Laplace equation (i.e.\ the equation satisfied by $U_0$ or
$U'_0$) becomes simply $\rmd ( \rho \star \rmd U_0 ) = 0$ outside the axis
of symmetry.  For the $\tilde{\gamma}$-Laplace equation, we
first note the following simple identity, valid for any pair of
functions $f_1$, $f_2$ on $\mathbb{K}$,
\begin{eqnarray*}
\rmd f_1 \wedge \star \rmd f_2 = - \star \rmd f_1 \wedge \rmd f_2 = 
- \left ( \rmd f_1, \rmd f_2 \right )_{\gamma} \rmd\rho \wedge \rmd z .
\end{eqnarray*}
Thus the second equation in (\ref{eq:1ernst0}) can be rewritten,
away from the axis of symmetry, as
\begin{eqnarray}
\rmd \left ( \rho \star \rmd \Omega'_0 \right )  - 4  \rho  \rmd \Omega'_0
\wedge \star \rmd U_0 = 0. 
\label{eqOmegaonK}
\end{eqnarray}
Our aim is to find suitable regular $\tilde{\gamma}$-harmonic
functions, i.e.\ suitable solutions of this equation on the domain
$\mathbb{K}^E$ corresponding to the exterior domain $\Do$.  More
precisely, the surface $\supo \subset \mathbb{E}^3$ projects into a
line $c_0$ in $\mathbb{K}$ defined parametrically as
$\{  z = z_0 (\mu), \rho = \rho_0 (\mu) \}$.
This line separates $\mathbb{K}$ into two
regions, the exterior (denoted by $\mathbb{K}^E$) and the interior.
The assumption we made on the topology of $\supo$, namely that it has
vanishing genus, implies that $\mathbb{K}^E$ is simply connected and
that $c_0$ intersects $\rho=0$ at two values of $z$, namely $z_S$ and
$z_N$.  In order to determine suitable solutions of
$\triangle_{\tilde{\gamma}} u = 0$ the following lemma is useful.
\begin{lemma}
  For any $\pw \in (z_S, z_N )$ and for any point $(\rho,z) \in
  \mathbb{K}^E$ define $\al(\rho,z) \in [0, 2\pi )$ by
\begin{eqnarray*}
\label{eq:alpha}
\cos \al(\rho,z) \equiv \frac{z-\pw}{\sqrt{\rho^2+(z-\pw)^2}},\\
\sin \al(\rho,z) \equiv \frac{\rho}{\sqrt{\rho^2+(z-\pw)^2}}.
\end{eqnarray*}
Then the PDE
\begin{equation}
\rmd Z_{\pw}=\cos \al \,\rmd U_0 +\sin \al \,\star \rmd U_0, 
\label{eq:W}
\end{equation}
with boundary condition $\lim_{\rho^2 + z^2 \rightarrow \infty}
Z_{\pw} = 0$ admits a unique solution on $\mathbb{K}^E$.
\label{equationW}
\end{lemma}
\textit{Proof:} The function $\al$ is regular everywhere in $\mathbb{K}$
except at $\{ z = \pw,\, \rho=0 \}$ which lies outside $\mathbb{K}^E$.
Similarly $U_0$ is regular everywhere on this simply connected domain.
So, in order to prove existence we only need to show that
$\rmd (\cos\al\,\rmd U_0 +\sin \al\,\star \rmd U_0 ) =0$ on $\mathbb{K}^E$.
A simple calculation which uses only the fact that $U_0$ is a
flat-harmonic function on $\Do$ gives
\begin{widetext}
\begin{eqnarray*}
\rmd (\cos\al\,\rmd U_0 +\sin \al\,\star \rmd U_0 ) = - \sin \al \left ( 
\rmd \al + \frac{1}{\tan \al} \star \rmd \al + \frac{1}{\rho} \rmd z \right
) \wedge \rmd U_0.
\end{eqnarray*}
\end{widetext}
Using the definition of $\al$ above it is immediate to show that the
term in parenthesis on the right hand side vanishes. Thus existence
of $Z_{\pw}$ follows. Furthermore, asymptotic flatness implies that
the right hand side of (\ref{eq:W}) tends to zero like $1/r^2$. Thus
$Z_{\pw}$ is bounded at infinity and we can impose boundary data
there. Since the general solution of (\ref{eq:W}) depends on an
arbitrary additive constant, the lemma follows.$\Box$

Note that both $\al$ and $U_0$ are analytic functions on $
\mathbb{K}^E$. It immediately follows that $Z_{\pw}$ is also analytic
on this domain.  Let us now define the function
\begin{eqnarray}
\Psi_{\pw} (\rho,z) =\frac{e^{ 2 U_0 - 2 Z_{\pw}}}{\sqrt{ \rho^2 + (z
    - \pw)^2}}, \qquad \pw \in (z_S, z_N). 
\label{principalOmega}
\end{eqnarray}
This function solves $\triangle_{\tilde{\gamma}} u = 0$, as we show next.
\begin{lemma}
  The function $\Psi_{\pw}$ defined in \textrm{(\ref{principalOmega})} is a
  regular $\tilde{\gamma}$-harmonic function on $\Do.$
\end{lemma}
\textit{Proof:} We need to show (i) $\triangle_{\tilde{\gamma}}
\Psi_{\pw} =0$ on $\Do$, (ii) $\Psi_{\pw}$ admits a $C^1$ extension to
$\partial \Do$ and (iii) $r \Psi_{\pw}$ is bounded at infinity.
Property (iii) is immediate from the corresponding property of
$Z_{\pw}$. Furthermore $U_0$ is $C^1$ on $\partial \Do \cup \Do$,
which implies, from the PDE (\ref{eq:W}), that $Z_{\pw}$ is also $C^1$
on this subset.  In addition $\al$ is analytic except at the point
$\{z = \pw,\, \rho =0 \}$; hence (ii) holds. In order to prove (i) it is
sufficient to check the differential equation
\begin{eqnarray*}
\rmd \left ( \rho \star \rmd \Psi_{\pw} \right ) - 4 \rho \rmd \Psi_{\pw}
\wedge \star \rmd U_0 = 0,
\end{eqnarray*}
which is equivalent to $\triangle_{\tilde{\gamma}} \Psi_{\pw}=0$
except on the symmetry axis $\rho=0$ ( $\triangle_{\tilde{\gamma}}
\Psi_{\pw}=0$ on the axis follows from the fact that $\Psi_{\pw}$ is
$C^2$ on $\Do$ -- in fact analytic).  
Note that $\Psi_{\pw}e^{-2 U_0 + 2 Z_{\pw}}$ is just the 
 flat-space harmonic function $1/r$ where $r$ is the distance from
 $\rho=0,~z=y$.
 Using this and $\rmd Z_{\pw}
\wedge \star \rmd Z_{\pw} = \rmd U_0 \wedge \star \rmd U_0$, a simple
calculation gives
the result.$\Box$

We can now state the necessary and sufficient conditions that the
boundary data must satisfy so that $\triangle_{\ovg} u = j$ admits a
decaying solution.

\begin{widetext}
\begin{theorem}
\label{Compatibility}
Let $f_0$, $f_1$ be continuous axially symmetric functions on a $C^1$
simply connected, axially symmetric surface $\supo$ of $\mathbb{E}^3$.
Let this surface be defined in cylindrical coordinates by $\{ \rho
= \rho_0 (\mu),\, z = z_0 (\mu),\, \phi = \varphi \}$, where $\mu$ takes
values in $[\mu_S,\,\mu_N ]$ and $\mu_S < \mu_N$ are the only solutions
of $\rho_0(\mu)=0$. Call $z_S \equiv z (\mu_S)$ and $z_N \equiv
z(\mu_N)$ and assume $z_S < z_N$ (i.e.\ that these values correspond to
the ``south'' and ``north'' poles of the surface, respectively).
Denote by $\Do$ the exterior region of this surface and let $j$ be any
axially symmetric function on $\Do$ such that $r^4 j$ is bounded at
infinity. Let $\gamma$ be the flat metric and $\tilde{\gamma} = e^{-8
  U_0} \gamma$, where $U_0$ is any regular flat-harmonic function on
$\D_0$.  Then the Cauchy problem
\begin{eqnarray*}
\triangle_{\ovg} u = j, \qquad u |_{\supo} = f_0, \qquad \vec{n} (u)
|_{\supo} = f_1, 
\end{eqnarray*}
where $\vec{n} |_{\supo} = - \dot{z}_0 \partial_{\rho} + \dot{\rho}_0
\partial_z$, admits a regular solution if and only if the
compatibility conditions
\begin{eqnarray}
\ovg = \gamma : &    
  \int_{\mu_S}^{\mu_N} \left . 
\left[\psi_{\pw} \, f_1 - f_0 \,\vec n(\psi_{\pw})
 \right]\rho_0 \right |_{\supo} \rmd \mu 
=  \frac{1}{2 \pi} \int_{\Do}\psi_{\pw} j
  \eta_{\gamma}, \qquad \forall \pw \in 
( z_S, z_N ) \label{forgamma}\\[4pt]
\ovg  =  \tilde{\gamma} : & 
  \int_{\mu_S}^{\mu_N} \left . 
\left[\Psi_{\pw} \, f_1 - f_0 \,\vec n(\Psi_{\pw})
 \right]\rho_0 e^{-4 U_0} \right |_{\supo} 
\rmd \mu  = \frac{1}{2 \pi} \int_{\Do}\Psi_{\pw} j e^{-12 U_0 }
  \eta_{\gamma}, \qquad \forall \pw \in 
( z_S, z_N ) \label{fortildegamma}
\end{eqnarray}
are satisfied, where $\psi_y$ and $\Psi_y$ are given in
(\ref{principalU}) and (\ref{principalOmega}) respectively.
\end{theorem}
\textit{Proof:} We give the proof for $\ovg = \tilde{\gamma}$; the case $\ovg
= \gamma$ follows by setting $U_0=0$ everywhere. Necessity follows
directly from Green's identity (\ref{Green}). In order to prove
sufficiency, let $u$ be the unique regular solution of the Dirichlet
problem $\triangle_{\tilde{\gamma}} u = j$, $u |_{\supo} = f_0$ on
$\Do$ (which is known to exist). Define $\vec{n} (u) |_{\supo}
= \tilde{f}_1$. Green's identity and the fact that $\Psi_{\pw}$ is a
regular $\tilde{\gamma}$-harmonic function implies
\begin{eqnarray*} 
 \int_{\mu_S}^{\mu_N} \left . 
\left[\Psi_{\pw} \, \tilde{f}_1 - f_0 \,\vec n(\Psi_{\pw})
 \right]\rho_0 e^{-4 U_0} \right |_{\supo} 
 \rmd \mu= \frac{1}{2 \pi} \int_{\Do}\Psi_{\pw} j e^{-12 U_0 }
  \eta_{\gamma}.
\end{eqnarray*}
Subtracting (\ref{fortildegamma}) we get $\int_{\mu_S}^{\mu_N}
\left . \Psi_{\pw}  ( f_1 - \tilde{f}_1 ) \rho_0 e^{-4 U_0} \right |_{\supo} 
\rmd \mu
=0$. We now apply Lemma \ref{integral} with the function $h
\equiv e^{- 2 U_0 - 2 Z_{\pw}} |_{\supo} \rho_0 (f_1 - \tilde{f}_1 )$.  It
follows that $h=0$ and hence $\tilde{f}_1 = f_1$. Thus the Cauchy problem
is solvable and the theorem follows.$\Box$
\end{widetext}

For the first order perturbations, the inhomogeneous term $j$ vanishes
and this theorem provides necessary and sufficient conditions
involving the boundary data only (and hence conditions on the interior
perturbations via the perturbed matching conditions described in the
previous section).

For the second order perturbations things are not so easy because the
equations are inhomogeneous ($j \neq 0$). In principle, one would need
to integrate the first order functions $U'_0$, $\Omega'_0$ in order to
compute $j$ and thence $\int_{\Do} \psi_{\pw} j \eta_{\gamma}$ (and the
corresponding expression for $\tilde{\gamma}$). In some practical
situations, an alternative procedure would be finding a particular
solution $u_p$ of
\[
\triangle_{\ovg} u=j.
\]
so that the homogeneous compatibility conditions (i.e.\ with $j=0$) can
be applied to the function $u_h=u-u_p$, which solves the homogeneous
equation $\triangle_{\ovg} u_h=0$. More specifically we should need to
check whether the Cauchy data $\{f_0-u_p|_{\supo},\, f_1 -\vec
n(u_p)|_{\supo}\}$ (where $f_0$ and $f_1$ are the Cauchy data for $u$)
satisfy the homogeneous compatibility conditions in theorem
\ref{Compatibility}.  Obviously this approach relies heavily on knowing a
particular solution of the inhomogeneous equation.

It is clear that, generically, we shall not be able to integrate the
first order equations explicitly or find a particular solution of the
inhomogeneous equation. We should still like to be able to treat the
problem in a satisfactory way.  The key idea which enables us to do so
is to rewrite the volume integrals on the right-hand sides of
(\ref{forgamma}) or (\ref{fortildegamma}) as surface integrals, and
hence rewrite everything in terms of boundary data (or integrals
thereof). Suppose that we were able to find an axially symmetric
vector $\vec{T} = T^{\rho} \partial_{\rho} + T^z \partial_z$ on $\Do$
such that (with $\nabla_a$ denoting covariant derivative with respect
to $\gamma$) $\nabla_a T^a = \psi_{\pw} j$ (for $\gamma$) or $\nabla_a
T^a = \Psi_{\pw} j e^{-12 U_0}$ (for $\tilde{\gamma}$). Then we could
use Gauss' identity to transform the volume integral into a surface
integral on $\supo$.  We now show that this is indeed possible. As
before, it is useful to work on the two-dimensional space
$\mathbb{K}^E$. Defining the one-form $\bm{T} \equiv T^{\rho} \rmd
\rho + T^z \rmd z$ we can translate the equation above for $\vec{T}$
into an equation for $\bm{T}$ on $\mathbb{K}^E$ by means of the
general identity $\rmd ( \rho \star \bm{T} ) = - \rho \nabla_a T^a
\rmd \rho \wedge \rmd z$.

We start with the equation for $U''_0$ in (\ref{eq:2ernst0}). We
obviously have $j = - e^{-4 U_0} ( \rmd \Omega'_0, \rmd \Omega'_0
)_{\gamma}$ so that the equation for $\bm{T}$ reads, using the
explicit form (\ref{principalU}) for $\psi_{\pw}$,
\begin{eqnarray}
\rmd \left (\rho \star \bm{T} \right ) + \sin \al e^{-4 U_0} \rmd \Omega'_0
\wedge \star \rmd \Omega'_0 = 0. 
\label{equationTU}
\end{eqnarray}
Our aim is to find a solution of this equation. In order to do this,
the following Lemma turns out to be useful.
\begin{lemma}
  Let $\Omega'_0$ be a solution of the second equation in
  (\ref{eq:1ernst0}) and $Z_{\pw}$ be defined as in (\ref{eq:W}). Then
  the equations
\begin{eqnarray}
\rmd \QQ_1 & =&  e^{-2 U_0 + 2 Z_{\pw} } \left [ - \left ( 1 + \cos \al
  \right ) \rmd \Omega'_0 - \sin \al \star \rmd \Omega'_0  
\right ], \nonumber\\[-5pt] \label{equationQ1}\\[-5pt]
\rmd \QQ_2 &  = & e^{-2  U_0 - 2 Z_{\pw} } \left [  \left ( 1 - \cos \al
  \right ) \rmd \Omega'_0 - \sin \al \star \rmd \Omega'_0 
\right ],  \nonumber
\end{eqnarray}
admit unique solutions which tend to zero at infinity. 
\label{Q12}
\end{lemma}
\textit{Proof:} The solutions, if they exist, are unique except for
additive constants. Asymptotic flatness of $\Omega'_0$ implies that
$\QQ_1$ and $\QQ_2$ each tend to a constant at infinity; these additive
constants can be chosen so that $\QQ_1$ and $\QQ_2$ vanish at infinity.
Thus the only non-trivial part of the proof is to show existence of
the solutions. Simple-connectedness of $\mathbb{K}^E$ implies that we
only need to check $\bm{\omega} \equiv \rmd ( e^{-2 U_0 + 2 \delta
  Z_{\pw} } [ - ( \delta + \cos \al ) \rmd \Omega'_0 - \sin \al \star \rmd
\Omega'_0 ]) = 0$, where $\delta = \pm 1$. A straightforward, if
somewhat long, calculation using the equations for $U_0$, $\Omega'_0$
and $Z_{\pw}$ gives
\begin{eqnarray*}
\bm{\omega} = e^{2 \delta Z_{\pw}-2 U_0 } \sin \al \left [ \rmd \al +
  \frac{\star \rmd \al }{\tan \al} + \frac{\rmd z}{\rho}
\right ] \wedge \rmd \Omega'_0 = 0,
\end{eqnarray*}
where again the explicit expression for $\al$ is used in the last
equality. $\Box$

The equations for $\QQ_1$ and $\QQ_2$ imply
$\rmd\QQ_1 \wedge \rmd\QQ_2 = 2 \sin \al
e^{-4 U_0} \rmd \Omega'_0 \wedge \star \rmd \Omega'_0$, which is the
crucial fact allowing us to solve (\ref{equationTU}). Indeed, defining
\begin{eqnarray}
\bm{T_1} \equiv \frac{1}{2 \rho} \QQ_1 \star \rmd \QQ_2, \label{T1}
\end{eqnarray}
it is immediate to check that (\ref{equationTU}) is satisfied for
$\bm{T} = \bm{T_1}$.  This expression is apparently singular at
$\rho=0$. However, on the axis of symmetry we have $\cos \al = + 1$ or
$\cos \al = -1$ depending on whether we are above the north pole or
below the south pole, respectively. Consequently, we have $\QQ_1 =0$ on
the subset of the symmetry axis below the south pole (from the
equation it satisfies and the fact that $\QQ_1$ vanishes at infinity)
and $\rmd\QQ_2$ vanishes on the axis above the north pole. Combining this
with analyticity of $\Omega'_0$ everywhere on $\Do$ (including the
axis), regularity of $\bm{T}_1$ follows.

Considering the equation for $\Omega''_0$,  in this case we have $j = 8
\left ( \rmd \Omega'_0, \rmd U'_0 \right )_{\tilde{\gamma}}$. Using the
explicit expression (\ref{principalOmega}) for $\Psi_{\pw}$, the
equation we need to satisfy is
\begin{eqnarray*}
\nabla_a T^a = \frac{8}{\sqrt{ \rho^2 + ( z - \pw )^2}} e^{-2 U_0 - 2
  Z_{\pw} } \left ( \rmd \Omega'_0, \rmd U'_0 \right )_{\gamma},
\end{eqnarray*}
which, in terms of exterior forms, reads
\begin{eqnarray}
\rmd \left ( \rho \star \bm{T} \right ) = 8 \sin \al e^{-2 U_0 - 2
  Z_{\pw} } \rmd \Omega'_0 \wedge \star \rmd U'_0. 
\label{equationTOmega}
\end{eqnarray}
We now use the fact that $U'_0$ is a flat-harmonic function. Thus
(compare Lemma \ref{equationW}) we can uniquely define a function
$Z'_{\pw}$ on $\mathbb{K}^E$ by the equation
\begin{eqnarray*}
\rmd Z'_{\pw}=\cos \al \,\rmd U'_0 +\sin \al \,\star \rmd U'_0, 
\qquad \lim_{\rho^2 + z^2 \rightarrow \infty} Z'_{\pw} = 0.
\end{eqnarray*}
It is then straightforward to check that
\begin{eqnarray}
\bm{T_2} \equiv  - \frac{4}{\rho} \QQ_2 \star \rmd \left ( Z'_{\pw}+ U'_0
\right) \label{T2} 
\end{eqnarray}
satisfies (\ref{equationTOmega}). Again, regularity at the axis
follows because $\QQ_2$ vanishes on the axis above the north pole and $d
(Z'_{\pw} + U'_0)$ is zero on the axis below the south pole.

We end this section by summarizing its main results in the form of the
following Theorem.
\begin{widetext}
\begin{theorem}
\label{casebycase}
Let the assumptions and notation of theorem \ref{Compatibility} hold.
Then

\noindent (i) the Cauchy boundary value problem
\begin{eqnarray*}
\triangle_{\gamma} U'_0 =0, \qquad U'_0 |_{\supo} = f_0, \qquad \vec{n}
\left (U'_0 \right ) |_{\supo} =f_1,
\end{eqnarray*} 
admits a regular solution on $\Do$ if and only if
\begin{eqnarray*}
 \int_{\mu_S}^{\mu_N} \left . 
\left[\psi_{\pw} \, f_1 - f_0 \,\vec n(\psi_{\pw})
 \right]\rho_0 \right |_{\supo} 
\rmd \mu = 0, \qquad \forall  \pw \in(z_S,\,z_N),
\end{eqnarray*}
(ii) the Cauchy boundary value problem
\begin{eqnarray*}
\triangle_{\gamma}\Omega'_0- 4\left(\rmd \Omega'_0,\rmd
  U_0\right)_\gamma=0, \qquad \Omega'_0|_{\supo} = f_0,
\qquad \vec{n} \left ( \Omega'_0 \right ) |_{\supo} = f_1,
\end{eqnarray*}
admits a regular solution on $\Do$ if and only if
\begin{eqnarray*}
 \int_{\mu_S}^{\mu_N} \left . 
\left[\Psi_{\pw} \, f_1 - f_0 \,\vec n(\Psi_{\pw})
 \right]\rho_0 e^{-4 U_0 } \right |_{\supo} 
\rmd \mu  = 0, \qquad \forall
 \pw \in (z_S,\,z_N), 
\end{eqnarray*}
(iii) the Cauchy boundary value problem
\begin{eqnarray*}
\triangle_{\gamma} U''_0+e^{-4U_0}
      \left(\rmd \Omega'_0,\rmd \Omega'_0\right)_\gamma=0, \qquad
      U''_0|_{\supo} = f_0, 
\qquad \vec{n} \left ( U''_0 \right ) |_{\supo} = f_1,
\end{eqnarray*}
admits a regular solution on $\Do$ if and only if
\begin{eqnarray*}
 \int_{\mu_S}^{\mu_N} \left . 
\left[\psi_{\pw} \, f_1 - f_0 \,\vec n(\psi_{\pw})
- \vec{n} \left ( \bm{T_1} \right )
 \right]\rho_0 \right |_{\supo}  \rmd \mu  = 0,
\qquad \forall  \pw \in (z_S,\,z_N),
\end{eqnarray*}
and (iv) the Cauchy boundary value problem
\begin{eqnarray*}
\triangle_{\gamma}\Omega''_0-8
      \left(\rmd \Omega'_0,\rmd U'_0\right)_\gamma
      -4\left(\rmd \Omega''_0,\rmd U_0\right)_\gamma=0,
\qquad \Omega''_0|_{\supo} = f_0,
\qquad \vec{n} \left ( \Omega''_0 \right ) |_{\supo} = f_1,
\end{eqnarray*}
admits a regular solution on $\Do$ if and only if
\begin{eqnarray*}
 \int_{\mu_S}^{\mu_N} \left . 
\left[ \left ( \Psi_{\pw} \, f_1 - f_0 \,\vec
 n(\Psi_{\pw}) \right ) e^{-4 U_0} 
 - \vec{n} \left ( \bm{T_2} \right )
 \right]\rho_0 \right |_{\supo} 
\rmd \mu  = 0, \qquad \forall  \pw \in (z_S,\,z_N),
\end{eqnarray*}
where $\psi_{\pw}$, $\Psi_{\pw}$, $\bm{T_1}$ and $\bm{T_2}$ are given
in (\ref{principalU}), (\ref{principalOmega}), (\ref{T1}) and
(\ref{T2}) respectively.
\end{theorem}
\end{widetext}

\textbf{Remark.} Since writing down $\bm{T_1}$ and $\bm{T_2}$ requires
the integration of $\rmd \QQ_2$ it may seem at first sight that one has not
really gained anything with respect to the volume integral in theorem
\ref{Compatibility}. The difference however is substantial because we only
need to know $\QQ_2$ \textit{on the boundary $\supo$}. So by projecting
equation (\ref{equationQ1}) into the boundary we get an ODE for $\QQ_2
|_{\supo}$ which can in principle be solved by quadratures. Hence the problem
reduces to performing integrals, which is of course much easier than
solving PDEs. Thus the compatibility conditions are truly written in
terms of the boundary data \textit{alone}, as we wished.

\section{Perturbations around spherically symmetric static
background configurations}
\label{sphere}

Everything we have discussed so far holds for any stationary and
axially symmetric perturbation of a static and axially symmetric
background. Of course in many cases of physical interest, and
especially for perfect fluids, one expects that equilibrium
(non-rotating) configurations of isolated bodies are spherically
symmetric. In the case of perfect fluids this has been rigorously
proven for a large class of equations of state \cite{BeiSim92}. It
is therefore of interest to specialize the previous results to the
case when the background spacetime is in fact spherically symmetric.
This implies in particular that the exterior background metric, being
vacuum, corresponds to the Schwarzschild metric
\begin{widetext}
\begin{eqnarray*}
\rmd s^2=-\left(1-\frac{2 m}{r}\right) \rmd t^2
+\left(1-\frac{2 m}{r}\right)^{-1}\rmd r^2+
r^2\left(\rmd \theta^2+\sin^2\theta\,\rmd \phi^2\right),
\end{eqnarray*}
\end{widetext}
where $m$ is the total mass of the spacetime.
The static Killing vector is given by $\vec{\xi}^{\,E}_0=\partial_t$,
and we fix the axial Killing vector to be $\vec{\eta}^{\,E}_0
= \partial_{\phi}$ in these coordinates. (The non-uniqueness in the
choice of axial symmetry in a spherically symmetric situation is
precisely what allows us to orient the coordinate system so that this
vector is the limit of the axial symmetries of the perturbed spaces.)
{}From the definition of $U_0$ we get
\begin{equation}
U_0=\frac{1}{2}\ln \left(1-\frac{2 m}{r}\right).\label{eq:u0}
\end{equation}
The intrinsic definition of the Weyl coordinate $\rho$,
$\rho^2 = - (\vec{\xi}^{\,E}_0 , \vec{\xi}^{\,E}_{0} )_{\gEo}
(\vec{\eta}^E_0 , \vec{\eta}^E_{0} )_{\gEo} + (\vec{\xi}^{\,E}_0 , 
\vec{\eta}^E_{0} )_{\gEo}^2,$ gives
\begin{equation}
\rho = r \sin \theta \sqrt{ 1 - \frac{2m}{r} }.
\label{eq:rho}
\end{equation}
In order to determine $z$ we use $(\rmd z,\rmd\rho)_{\gEo}
=0$, $(\rmd z,\rmd z)_{\gEo} = (\rmd\rho,\rmd\rho)_{\gEo}$, which can be
solved to give $z = z_0 \pm (r-m) \cos \theta$, where $z_0$ is a
constant. With these coordinate changes, $\{r,\,\theta,\,\phi\}$ can be
regarded as coordinates on $(\mathbb{E}^3,\,\gamma)$.  Choosing $\{\rmd
r,\,\rmd\theta\}$ to be positively oriented on a plane $\phi
= \mbox{const.}$, i.e.\ that $\star \rmd r \propto \rmd \theta$ with a
positive proportionality factor, and choosing
(as we have above) that $\{ \rmd z,\, \rmd \rho \}$ is also positively
oriented, the $+$ sign above is selected.  We can further choose
$z_0=0$ without loss of generality, so we have
\begin{equation}
z = \left (r - m \right ) \cos \theta.
\label{eq:z}
\end{equation}
Spherical symmetry of the whole background spacetime requires that the
surface of the body $\supEo$ is defined by $r=\rs (>2m)$. The embedding
for this surface can be chosen to be $\embed^E_0 : \{ \tau,\,\varphi,\,\mu
\} \rightarrow \{t=\tau,\, \phi=\varphi,\, r=\rs,\, \theta=\pi-\mu\}$, with
$\varphi\in[0,\,2\pi)$ and $\mu\in[0,\,\pi]$. This choice for $\mu$ is
motivated by the fact that it increases from the south to
the north pole of the body, as required. In fact, we have $\mu_S=0$
and $\mu_N=\pi$. The range for the constant $\pw$ introduced in the
previous section is then given by $-\rs+m<\pw<\rs-m$.  With this
embedding we clearly have
\begin{equation}
  \label{eq:supo}
  \rho_0 (\mu)=\rs\sin \mu\sqrt{1-\frac{2 m}{\rs}},~~~~
  z_0 (\mu) =-(\rs-m) \cos\mu
\end{equation}
The vector $\vec{e}_0 = \rmd \embed^E_0 (\partial_\mu )$ has norm
$r_0^2$, which implies that $\vec{n}_0$ (which is defined to have the
same norm) reads
\begin{equation}
\label{eq:normal}
\vec n_0= \left. -\rs \sqrt{1-\frac{2 m}{\rs}}\,\partial_r \right |_{\supo}.
\end{equation}
We can now write down the first and second order perturbed Ernst
equations in Schwarzschild coordinates. Performing the coordinate
transformation (\ref{eq:rho}), (\ref{eq:z}) the flat metric $\gamma$
in $\mathbb{E}^3$ becomes
\begin{widetext}
\begin{eqnarray*}
\gamma = \frac{ \left (r-m \right )^2 - m^2 \cos^2 \theta}{r^2} \left ( 
\frac{dr^2}{1 - \frac{2m}{r}}
+ r^2 \rmd \theta^2 \right ) + r \left (r- 2m \right ) \sin^2 \theta
\rmd \phi^2,
\end{eqnarray*}
which implies the following form for the first order Ernst equations
(\ref{eq:1ernst0})
\begin{eqnarray}
r \left (r-2m \right ) 
U'_0{}_{,rr}+2(r-m)U'_0{}_{,r}+U'_0{}_{,\theta\theta}+
\frac{\cos\theta}{\sin \theta}U'_0{}_{,\theta} = 0, \label{eq:equp}
\nonumber\\
r \left( r-2m \right) \Omega'_0{}_{,rr} + 2(r-3m)\Omega'_0{}_{,r}+
\Omega'_0{}_{,\theta\theta}+\frac{\cos\theta}{\sin\theta}
\Omega'_0{}_{,\theta}=0. \label{eq:omegap}
\end{eqnarray}
The second order equations (\ref{eq:2ernst0}) become
\begin{eqnarray*}
r \left (r-2m \right ) 
U''_0{}_{,rr}+2(r-m)U''_0{}_{,r}+U''_0{}_{,\theta\theta}+
\frac{\cos\theta}{\sin \theta}U''_0{}_{,\theta}
+ \frac{r^3  \Omega'_0{}_{,r}^2}{r- 2m}
+\frac{r^2  \Omega'_0{}_{,\theta}^2 }{\left ( r- 2m \right )^2}
= 0, \label{eq:equpp} \\
r \left( r-2m \right) (\Omega''_0{}_{,rr} -8\Omega'_0{}_{,r}
U'_0{}_{,r})
 + 2(r-3m)\Omega''_0{}_{,r}+
\Omega''_0{}_{,\theta\theta}+\frac{\cos\theta}{\sin\theta}
\Omega''_0{}_{,\theta}
- 8 \Omega'_0{}_{,\theta} U'_0{}_{,\theta}  =0. \label{eq:omegapp}
\end{eqnarray*}
\end{widetext}
Let us analyze the equation for $U'_0$. 
Expanding in axisymmetric spherical harmonics
\[
U'_0 = \sum_{l=0}^{\infty} u^{(1)}_l(r) P_l(\cos\theta),
\]
we get a collection of ordinary differential equations which, after 
performing the change of variables
$r = m (1+x)$, read
\[
\left ( 1- x^2 \right ) u^{(1)}_{l,xx} - 2 x u^{(1)}_{l,x} + l \left (l+1 
\right ) u^{(1)}_l =0.\]
These equations can be solved as 
$u^{(1)}_l = a_l P_l(x) + b_l P_l(x) 
\int_{+ \infty}^x \frac{dy}{\left ( 1- y^2 \right ) P_l^2(y)}$, 
where $P_l(x)$ is the $l$-th Legendre polynomial and $a_l$ and $b_l$
are constants (note that
the integral in the second summand converges). 
Asymptotic flatness demands that $U'_0$ tends to zero when $x \rightarrow 
 \infty$. Thus $a_l=0$ for all $l$ and the most general solution for $U'_0$ is 
\[
U'_0 = \sum_{l=0}^{\infty} b_l P_l(\cos\theta)
P_l \left (\frac{r}{m}-1 \right ) 
\int_{+ \infty}^{\frac{r}{m}-1} \frac{dy}{\left ( 1- y^2 \right ) P_l^2(y)}, 
\]
This expression also gives the general solution of the homogeneous part of 
$U''_0$. For $\Omega'_0$, the expansion in spherical harmonics
$\Omega'_0=\sum_{l=0}^{\infty} w^{(1)}_l(r) P_l(\cos\theta)$
transforms (\ref{eq:omegap}) into
\[
\left ( 1- x^2 \right ) w^{(1)}_{l,xx} - 2 \left ( x - 2 \right )  
w^{(1)}_{l,x} + l \left (l+1 \right ) w^{(1)}_l =0
\]
where again the $x$ variable has been used. 
The general solution is now
\begin{widetext}
\noindent
$w^{(1)}_l = c_l P^{(-2,2)}_l(x) + d_l P^{(-2,2)}_l(x) 
\int_{+ \infty}^x \frac{(y-1) dy}{\left ( y+1 \right )^3 (
  P^{(-2,2)}_l(y) )^2},$
where $P_l^{(-2,2)}(x)$ is the Jacobi polynomial and $c_l$ and $d_l$
  are constants (we are using the
notation of \cite{GraRyz65}). Imposing the condition that the
  solution tends to 
zero at infinity we conclude that $c_l=0$ and we can write the general
solution as 
\[
\Omega'_0 = \sum_{l=0}^{\infty} d_l P_l(\cos\theta)
P^{(-2,2)}_l \left (\frac{r}{m}-1 \right ) 
\int_{+ \infty}^{\frac{r}{m}-1} \frac{(y-1) dy}{\left ( y+1 \right )^3
  ( P^{(-2,2)}_l(y) )^2}.
\]
\end{widetext}
Having obtained these solutions, we consider next the boundary
data that these functions must satisfy. In Proposition 
\ref{FullBoundary} we obtained expressions that involve only tangential 
derivatives of several scalar objects. We need to evaluate them.
First of all we notice
that spherical symmetry implies that
\begin{equation}
  \label{eq:tanUes0}
  \tang(U_0|_{\supo})=\tang(\vec n(U_0)|_{\supo})=0,
\end{equation}
which substantially simplifies the boundary conditions. The explicit
forms of $\vec{n} (U_0) |_{\supo}$, $\X_0$ and $\X_1$ are directly
obtained from their definitions as
\begin{eqnarray*}
\vec{n} (U_0 ) |_{\supo} &=& \frac{-1}{\sqrt{\xs^2 -1 }}, \\
\X_0 &=& - \frac{\xs \sqrt{\xs^2 -1 }}{\xs^2 - \cos^2 \mu},
\qquad \X_1 = \frac{\cos \mu \sin \mu}{\xs^2 - \cos^2 \mu},
\end{eqnarray*}
where $\xs$ is the value of $x$ on the surface, i.e.\ $\xs = \rs/m -1$.
Since the background spacetime satisfies the matching conditions, we
obviously must have $\Usupo= 1/2\log ( (\xs -1)/(\xs+1))$ and
$\nUsupo= -1/\sqrt{\xs^2-1}$.  The functions $\P_1$, $\Q_1$,
$\P_2$ and $\Q_2$ are still arbitrary because they determine how the
unperturbed surface is deformed to first and second order. The same is
true regarding the functions $\Usupprime$, $\nUsupprime$,
$\Usupprimeprime$, $\nUsupprimeprime$ for the $U$-boundary conditions
and $\Omegasupprime$, $\nOmegasupprime$, $\Omegasupprimeprime$,
$\nOmegasupprimeprime$ for the $\Omega$-boundary conditions, which
depend on the interior perturbations.  Using all these expressions,
the boundary conditions described in Proposition \ref{FullBoundary}
simplify, for spherical backgrounds, to
\begin{widetext}
\begin{eqnarray*}
\Uprimeo & = &  \Usupprime  + \Q_1 \minusnUSch , \qquad
\nUprimeo  =   \nUsupprime + \left ( \frac{d \P_1  }{d \mu}
 + \Q_1 \frac{\xs}{\sqrt{\xs^2-1}} \right ) \minusnUSch  , \\
\Uprimeprimeo & = &  \Usupprimeprime - 2 \P_1 \frac{d \Usupprime}{d \mu } - 
 2 \Q_1  \nUsupprime-
 \left ( 2 \frac{d (\P_1 \Q_1) }{d \mu} - \Q_2  + \Q_1^2
 \frac{\xs}{\sqrt{\xs^2-1}} \right . \\ 
& &  \left .
 + \frac{ \left ( \P_1^2 - \Q_1^2 \right ) \xs \sqrt{\xs^2 -1 }
 +2 \P_1 \Q_1 \cos \mu \sin \mu}{\xs^2 - \cos^2 \mu} \right )
 \minusnUSch  \\
\nUprimeprimeo & = &  \nUsupprimeprime + 2 \frac{d}{d \mu} \left ( \Q_1 
 \frac{d \Usupprime}{d \mu} \right )
 - 2 \frac{d \left ( \P_1 \nUsupprime \right ) }{d \mu}
 + 2 \Q_1 \left ( \frac{\cos \mu}{\sin \mu}
 \frac{d \Usupprime}{d \mu } - \frac{\xs}{\sqrt{\xs^2-1}} \nUsupprime \right )
\\
& &
 - \left [ \frac{d^2}{d \mu^2} \left (  \P_1^2 - \Q_1^2  \right )
 + \frac{d}{d \mu} \left (
 - \P_2 + \frac{ \left ( \P_1^2  - \Q_1^2 \right ) \cos \mu \sin
 \mu}{\xs^2 -
   \cos^2 \mu}
 + 2\P_1\Q_1 \frac{\xs}{\sqrt{\xs^2-1}}\right )
  \right .\\
& &
 \phantom{2\P_1\Q_1}
 \left . -\frac{1}{\sin \mu}
  \frac{d}{d\mu} \left (\frac{2 \P_1 \Q_1 \xs \sqrt{\xs^2 -1 }\sin
  \mu}{\xs^2 -
   \cos^2 \mu}
   +\Q_1^2 \cos \mu) \right ) \right ] \minusnUSch \\
& &
 + \left (  \Q_2  - \frac{ (\P_1^2-\Q_1^2) 
 \xs \sqrt{\xs^2 -1 }}{\xs^2 - \cos^2 \mu}
 - 2 \P_1 \Q_1 \frac{\cos \mu}{ \sin \mu} \right )  \frac{\xs}{\xs^2-1}
\\ & & \\
\Omegaprimeo & =& \Omegasupprime, \quad \nOmegaprimeo= \nOmegasupprime, 
\quad
\Omega''_0 |_{\supo} = \Omegasupprimeprime-2 \P_1 \frac{d \Omegasupprime}{d 
\mu}
- 2 \Q_1 \nOmegasupprime, \\
\vec{n} \left ( \Omega''_0 \right )  |_{\supo} & = & \nOmegasupprimeprime +
2 \frac{d}{d \mu} \left ( \Q_1 \frac{d\Omegasupprime}{d \mu} \right )
-2 \frac{d\left ( P_1 \nOmegasupprime \right ) }{d \mu}
+ 2 \Q_1 \left [  \frac{\cos \mu}{\sin \mu} \frac{d \Omegasupprime}{d \mu}  
- \frac{\xs - 4}{\sqrt{\xs^2-1}}
\nOmegasupprime \right ].
\end{eqnarray*}
\end{widetext}

Our next aim is to find what compatibility conditions these data must
satisfy in order to admit asymptotically flat solutions.  In
particular we need to find the functions $\psi_{\pw}$ and $\Psi_{\pw}$
defined in the previous section.  Recalling the definition $\psi_\pw =
1/\sqrt{\rho^2 + (z-\pw )^2 }$ and using (\ref{eq:rho}) and
(\ref{eq:z}) we get the explicit form
\begin{eqnarray*}
\psi_\pw = \frac{1}{\sqrt{m^2 x^2 + \pw^2 - 2 m x \pw \cos \theta  -m^2 \sin^2 
\theta}},
\end{eqnarray*}
where $\pw$ is a constant satisfying $|\pw| \leq m x_0$.  For
$\Psi_{\pw}$ we first need to integrate the partial differential
equation (\ref{eq:W}) for $Z_{\pw}$.  Contracting this equation with
$\partial_{\theta}$ and using the fact that $U_0$ depends only on $r$
we get
\begin{eqnarray*}
\frac{ \partial Z_{\pw}}{\partial \theta} = 
\frac{m \sin \theta}{\sqrt{m^2 x^2 + \pw^2 - 2 m x \pw \cos \theta  -m^2 
\sin^2 \theta}},
\end{eqnarray*}
which, together with the boundary condition at infinity, gives
\begin{widetext}
\begin{eqnarray*}
  e^{-Z_{\pw}} = \frac{Z_0(x)}{\left ( \pw - m \right ) \sqrt{x^2-1}  }
\left [ \pw x  - m \cos{\theta} -
\sqrt{m^2 x^2 + \pw^2 - 2 m x \pw \cos \theta  -m^2 \sin^2 \theta} \right ]
\end{eqnarray*}
where $Z_0$ is an arbitrary function of $x$ obeying
$Z_0 \rightarrow
1$ as $x \rightarrow \infty$. 
(Using L'H\^opital's rule one can easily
check that this form for $e^{-Z_{\pw}}$
has the correct limit at the
special value $y=m$.) Then contracting (\ref{eq:W}) with $\partial_r$
we find that $Z_0$ is constant. Thus
\begin{eqnarray*}
  e^{-Z_{\pw}} = \frac{\pw x  - m \cos{\theta} -
\sqrt{ \frac{}{}m^2 x^2 + \pw^2 - 2 m x \pw \cos \theta  -m^2 \sin^2 \theta} 
}{\left ( \pw - m \right ) \sqrt{x^2-1}  }.
\end{eqnarray*}
We can finally write down the explicit expression for $\Psi_y$,
which reads
\begin{eqnarray*}
\Psi_{\pw} = 
\frac{ \left ( y x - m \cos \theta - \sqrt{ 
\frac{}{}m^2 x^2 + \pw^2 - 2 m x \pw \cos \theta  -m^2 \sin^2 \theta }
  \right )^2}{\left (y-m \right )^2 \left (x+1 \right )^2 
\sqrt{\frac{}{}m^2 x^2 + \pw^2 - 2 m x \pw \cos \theta  -m^2 \sin^2 \theta}}.
\end{eqnarray*}
\end{widetext}
Having obtained $\psi_{\pw}$ and $\Psi_{\pw}$ explicitly, the
compatibility conditions that the boundary data for $U'_0$,
$\Omega'_0$, $U''_0$ and $\Omega''_0$ must satisfy in the spherically
symmetric case are direct consequences of theorem \ref{casebycase} in
the previous section.

\begin{acknowledgments}
This research was primarily funded by EPSRC grant GR/R 53685, which
supported R. Vera as a Research Assistant and M. Mars as a Visiting
Fellow.  We are grateful to the other Visiting Fellow supported by the
grant, Prof.\ J.M.M. Senovilla, for helpful discussions. During the
final stages of preparation, R. Vera was supported by the Irish
Research Council for Science, Engineering and Technology postdoctoral
fellowship PD/2002/108, and later by a Basque Government grant
Ref. BFI05.335. M. Mars wishes to thank Robert Bartnik for pointing out
the connection of the problem with the Dirichlet-Neumann map and for
providing useful references. He also acknowledges financial support
from the Spanish Government under Project BFME2003-02121.  Many of the
calculations were verified with the help of the computer algebra
system REDUCE.
\end{acknowledgments}


\appendix*
\section{The perturbed boundary conditions}

In section \ref{sec:second} we introduced four functions $\Usupe$,
$\nUsupe$, $\Omegaesupe$ and $\nOmegaesupe$ such that the full matching
conditions for the Ernst potential are $\U |_{\supe} = \Usupe$,
$\vec{n}_{\ee} (\U) |_{\supe} = \nUsupe$, $\Omegae |_{\supe} = \Omegaesupe$,
$\vec{n}_{\ee} (\Omegae) |_{\supe} = \nOmegaesupe$. We
want to take first and second derivatives with respect to $\ee$ (at
fixed $\mu$) in order to obtain the perturbed matching conditions in
Proposition \ref{FullBoundary}. Since the calculations for $\U$ and
$\Omegae$ are very similar, let us introduce a function $\Fe(\rho,\,z,\,\ee)$
which satisfies
\begin{eqnarray}
\Fe |_{\supe} = \Fsupe, \qquad \vec{n}_{\ee}( \Fe) |_{\supe} = \nFsupe, 
\label{MFullForF}
\end{eqnarray}
for some functions $\Fsupe(\mu,\,\ee)$, $\nFsupe(\mu,\,\ee)$.  We
shall obtain the perturbed expressions for this generic function and
then specialize to $\U$ and to $\Omegae$. To do so we need to
differentiate (\ref{MFullForF}) with respect to $\ee$. The right hand sides
are defined as functions of $(\mu,\,\ee)$ so that for
them $d/d\ee~\equiv~\p/\p\ee$, but on the left hand sides we have
functions such as $\Fe(\rho,\,z,\,\ee)|_{\supe}
= \Fe(\rho(\mu,\,\ee),\,z(\mu,\,\ee),\,\ee)$. For the latter we have to
take a convective derivative with velocity $\vec{Z}_{1,\ee}$,
defined by the obvious
generalization of (\ref{zdef}) when we evaluate at $\supe$ instead
of $\supo$. $\vec{Z}_{2,\ee}$ is defined similarly.

Evaluating (\ref{MFullForF}) at $\epsilon=0$, we immediately get the
unperturbed boundary data $F_0 |_{\supo} = \Fsupo$, $\vec{n} (F_0)
|_{\supo} = \nFsupo$, and taking the first convective derivative of $\Fe$
gives (on simple rearrangement)
\[
\fper = \Fsupprime - \P_1 \efo - \Q_1 \nfo.
\]
To obtain the first derivative of $\vec{n}_{\ee}( \Fe) = n^i_{\ee}\p_i
\Fe$ we calculate
\begin{eqnarray*}
\left . \frac{d}{d \epsilon}(\vec{n}_{\ee}( \Fe)) \right |_{\supe} &=& 
 \left . \frac{\partial  n^i_{\ee}}{\partial \ee} \p_i \Fe + n^i_{\ee} \p_i
 \frac{d\Fe}{d \ee} \right |_{\supe} \\ 
 & = & \left .
\frac{\partial n^i_{\ee}}{\partial \ee} \p_i \Fe + Z^i_{1,\ee} n^j_{\ee}
 \p_i \p_j \Fe + n^i_{\ee} \p_i \Fe^\prime \right |_{\supe}.
\end{eqnarray*}
Substituting for $\vec{Z}_1$ and
$\frac{\p \vec{n}_\ee}{\p\ee}$ from
(\ref{zdef}) and (\ref{dndef}),
using the identity
\begin{equation}
e^i n^j \p_i \p_j \F_0|_{\supo}
= \partial_\mu (\nfo) - (X_0 \p_\mu (\fo)
 + X_1 \nfo) ,\label{enF}
\end{equation}
which like subsequent similar identities follows from
integration by parts using $\p_\mu(\vec{e})=X_1\vec{e}-X_0\vec{n}$ and
$\p_\mu(\vec{n})=X_0\vec{e}+X_1\vec{n}$, and rearranging, we obtain,
at $\ee=0$,
\begin{widetext}
\begin{eqnarray}
\label{Mnfper}
\nfper &=& \nFsupprime - \P_1 \enfo - \Q_1 \nnfo 
\\&&
+ 
\left ( \frac{ d \Q_1}{d \mu} + \Q_1 \X_1 \right ) \efo -
\left ( \frac{d \P_1}{d \mu} + \Q_1 \X_0 \right ) \nfo.\nonumber
\end{eqnarray}
\end{widetext}

The second derivative of $\Fe$ follows by applying the convective
derivative twice. Thus we obtain
\begin{eqnarray*}
\H''_{\ee} &=& (\Fe' + Z^i_{1,\ee} \p_i\Fe)'
 + Z^i_{1,\ee}\p_i(\Fe' + Z^j_{1,\ee} \p_j\Fe)\\
&=& \Fe'' +2Z^i_{1,\ee} \p_i \Fe'+ Z^i_{2,\ee} \p_i \Fe +Z^i_{1,\ee}Z^j_{1,\ee}
\p_i\p_j\Fe.
\end{eqnarray*}
Substituting the values of $\vec{Z}_1$ and $\vec{Z}_2$ from
(\ref{zdef}), using the identities (\ref{enF}) and
\begin{equation}
e^i e^j \p_i \p_j \F_0|_{\supo} = \eefo  -
(X_1 \efo  - X_0 \nfo ) ,\label{eeF}
\end{equation}
and rearranging we obtain
\begin{widetext}
\begin{eqnarray}
\label{Mfperper}
\fperper
 |_{\supo} &=&   \Fsupprimeprime - 2 \P_1 \efper|_{\supo} - 2 \Q_1 \nfper
- \Q_1^2 \nnfo
\nonumber \\ & &
- 2 \P_1 \Q_1 \enfo - \P_1^2 \eefo 
\nonumber \\ & &
- \left ( \Q_2 + \P_1^2 \X_0 - 2 \P_1 
\Q_1 \X_1 \right ) \nfo 
- \left ( \P_2 - \P_1^2 \X_1 - 2 \P_1 \Q_1 \X_0  \right ) \efo, \nonumber
\end{eqnarray}
The last of these expressions that we need is the second derivative of 
$\vec{n}_{\ee}( \Fe)$. This we can obtain by calculating
\begin{eqnarray*}
\left . [\vec{n}_{\ee}( \Fe)]'' \right |_{\supe}
&=&  \frac{\p^2 n^i_\ee}{\p\ee^2}\p_i \Fe +
  2\frac{\p n^i_\ee}{\p\ee}(\p_i \Fe'+ Z^j_{1,\ee}\p_i\p_j \Fe)
\\
 &&
\left  .
  +n^i_\ee(Z_{2,\ee}^j\p_i\p_j \Fe +Z^j_{1,\ee}Z^k_{1,\ee}\p_i\p_j\p_k \Fe
+ 2Z^j_{1,\ee}
  \p_j\p_i\Fe' + \p_i \Fe'') \right |_{\supe} 
\end{eqnarray*}
The terms in $\Fe''$ and $\Fe'$ on the right at $\ee=0$
are readily evaluated on
substituting for $\frac{\p \vec{n}_\ee}{\p\ee}$ and  $\vec{Z}_1$ from
(\ref{dndef}) and (\ref{zdef}), and
again using the identities (\ref{enF}) and (\ref{eeF}). They lead to
$\nfperper$ and
\[
 2 \Q_1 \nnfper + 2 \p_\mu(\P_1 \nfper)
- \left ( 2 \frac{d \Q_1}{d \mu} + 2 \Q_1 \X_1 \right )
\efper|_{\supo}
 + 2 \Q_1 \X_0 \nfper
\]
respectively. To evaluate $n^iZ^j_1Z^k_1\p_i\p_j\p_k \F_0 |_{\supo}$
in the form
we require, we will need the identities
\begin{eqnarray}
e^ie^jn^k\p_i\p_j\p_k \F_0|_{\supo} &=&
\p_{\mu\mu}(\nfo)-X_0\efo
  -X_1 \p_\mu(\nfo) - \dot{X}_0 \efo
    \label{eenF}\\
&&   -\dot{X}_1 \nfo -\X_0 e^i e^j \p_i \p_j \F_0|_{\supo} -
2\X_1 e^i n^j \p_i \p_j \F_0|_{\supo}
    +X_0n^in^j\p_i\p_j \F_0|_{\supo} , \nonumber \\ 
e^in^jn^k\p_i\p_j\p_k \F_0|_{\supo} &=&
\p_\mu( n^in^j\p_i\p_j\F_0|_{\supo}) -
   2X_0\enfo  -2X_1(n^in^j\p_i\p_j\F_0|_{\supo}) \label{nnnF}\\
&&  +2X_0(X_0\efo +X_1 \nfo). \nonumber
\end{eqnarray}
Using these, the previous identities (\ref{enF}) and (\ref{eeF}), and
the values of $\vec{Z}_1$, $\vec{Z}_2$, $\frac{\p \vec{n}_\ee}{\p\ee}$
and $\frac{\p^2 \vec{n}_\ee}{\p\ee^2}$ from (\ref{zdef}) and
(\ref{dndef}), we obtain
\begin{eqnarray}
\nfperper & = & \nFsupprimeprime - 2 \Q_1 \nnfper -2 \p_\mu(\P_1
  \nfper) 
 \nonumber \\
&& 
+ \left ( 2 \frac{d \Q_1}{d \mu} + 2 \Q_1 \X_1 \right ) \efper|_{\supo}
 - 2 \Q_1 \X_0 \nfper \nonumber \\
& &
  - \Q_1^2 \nnnfo -2 \P_1 \Q_1 \ennfo - \P_1^2 \eenfo \nonumber \\
& &
 - \left ( \Q_2 + 2 \Q_1^2 \X_0  + 2 \Q_1 \frac{d \P_1}{d \mu }  + \P_1^2 
 \X_0 - 2 \P_ 1 \Q_ 1 \X_1
 \right ) \nnfo \label{MAllFo} 
 \\
& &
 + \left ( - \P_2 + \frac{d (\Q_1^2-\P_1^2)}{d \mu} 
 + \P_1^2 \X_1 + 2 \Q_1^2 \X_1 \right ) \enfo
  \nonumber \\
 & &
 + \left ( 2 \P_1 \frac{d \Q_1}{d \mu}
 + 2 \P_1 \Q_1 \X_1 \right ) \eefo   \nonumber \\ & &
 +\left [ - \frac{d \P_2}{d \mu} + \frac{d (\P_1^2\X_1)}{d \mu}
 - \X_1 \frac{d (\Q_1^2)}{d \mu}+ 2 \P_1 \X_0 \frac{d \Q_1}{d \mu}
 \right . \nonumber \\
& & \phantom{+[} \left .
- \Q_2 \X_0  -\P_1^2\X_0^2 + 2 \P_1 \Q_1 \X_0 
\X_1 - 2
\Q_1^2 \X_1^2 \right ] \nfo  \nonumber \\ & &
+\left [  \frac{d \Q_2}{d \mu}
+  \frac{d (\P_1^2\X_0)}{d \mu}
- X_0\frac{d (\Q_1^2)}{d \mu} - 2  \P_1 \X_1\frac{d \Q_1}{d \mu}
\right. \nonumber \\
& & \phantom{+[}\left  .
 + \Q_2 \X_1  + 
\P_1^2 \X_0 \X_1 - 2 \P_1 \Q_1 
\X_1^2 - 2
\Q_1^2 \X_0 \X_1 \right ] \efo. \nonumber
\end{eqnarray}
\end{widetext}

These expressions are in fact identities which hold for any
axially symmetric function, in particular for $\U$ and $\Omegae$, but they are
not in a form given by the boundary data, since they involve $\nnfo$
and $\nnnfo$.  However, because the functions $U_0$, $U'_0$,
$\Omega'_0$ satisfy the elliptic equations (\ref{eq:ernst0}) and
(\ref{eq:1ernst0}) which relate second order tangential to second
order normal derivatives we can eliminate the second and third
order transverse derivatives. Since $\gamma^{ij}
|_{\supo} = \frac{1}{\sss} ( n^i n^j + e^i e^j ) + \frac{1}{\rho_0^2}
( \partial_\phi)^i ( \partial_\phi)^j |_{\supo} $, $\sss(\p_{\rho\rho}
  +\p_{zz}) F= (n^in^j+ e^i e^j )\p_i\p_j F$ and we immediately find
the identity
\begin{widetext}
\begin{equation}
\nnfo \equiv \sss (\Delta_{\gamma} F_0) |_{\supo} - \eefo + \X_1 \efo - \X_0 
\nfo -  \frac{\dot{\rho}_0}{\rho_0} \efo + \frac{\dot{z}_0}{\rho_0} \nfo.
\end{equation}
Applying this identity to $U_0$, $U'_0$ and $\Omega'_0$ and using the
field equations they satisfy, we obtain the following expressions for the
transverse-transverse derivatives on the boundary
\begin{eqnarray}
\nnUo & =  &  - \eeUo + \left ( \X_1 - \frac{\dot{\rho}_0}{\rho_0} \right )
\eUo
- \left ( \X_0 - \frac{\dot{z}_0}{\rho_0}  \right)
\nUo,  \label{MnnUo} \\
\nnUprimeo & =  &  - \eeUprimeo + \left ( \X_1 - \frac{\dot{\rho}_0}{\rho_0} 
\right )
\eUprimeo - \left ( \X_0 -\frac{\dot{z}_0}{\rho_0}  \right)
\nUprimeo, \label{MnnUprime} \\
\nnOmegaprimeo & =  &   4 \nOmegaprimeo \nUo + 4 \eOmegaprimeo \eUo
- \eeOmegaprimeo + \nonumber \\
& & + \left ( \X_1 - \frac{\dot{\rho}_0}{\rho_0} \right )
\eOmegaprimeo- \left ( \X_0 -\frac{\dot{z}_0}{\rho_0}  \right)  
\nOmegaprimeo. \label{MnnOmegaprime}
\end{eqnarray}
It only remains to evaluate $\nnnUo$ in terms of known boundary data.
In order to do that we need a similar identity but now
involving third derivatives. A straightforward calculation gives
\begin{eqnarray*}
\nnnfo & \equiv &  \sss \vec{n} \left ( \Delta_{\gamma} F_0 \right ) 
|_{\supo} - \eenfo +
2 \X_0  \eefo \nonumber \\
& & + \left ( 3 \X_1 - \frac{\dot{\rho}_0}{\rho_0} \right )\enfo
+  \left ( - \X_0 + \frac{\dot{z}_0}{\rho_0} \right )\nnfo \nonumber \\
& & + \left (\dot{\X}_0 - 3 \X_0 \X_1 + \frac{\dot{\rho}_0}{\rho_0} \X_0
- \frac{\dot{\rho}_0 \dot{z}_0}{\rho_0^2}
\right )  \efo \nonumber \\
& & + \left ( \dot{\X}_1 - 2 \X_1^2 + \X_0^2 +
\frac{\dot{\rho}_0}{\rho_0} \X_1 + \frac{\dot{z}_0^2}{\rho_0^2} \right ) \nfo,
\end{eqnarray*}
which, again, holds for any axially symmetric function in $\mathbb{E}^3$.
Applying it to $U_0$, and using $\Delta_{\gamma} U_0 =0$ and (\ref{MnnUo}), 
we obtain
\begin{eqnarray}
\nnnUo & \equiv &    - \p_{\mu\mu}(\nUo)
+ \eeUo \left ( 3 \X_0 - \frac{\dot{z}_0}{\rho_0} \right ) + \nonumber \\
& & + \enUo \left ( 3 \X_1 - \frac{\dot{\rho}_0}{\rho_0} \right ) + 
\nonumber \\
& & + \eUo \left (\dot{\X}_0 -
4 \X_0 \X_1 + 2 \frac{\dot{\rho}_0}{\rho_0} \X_0
+ \frac{\dot{z}_0}{\rho_0} \X_1 -  \frac{2 \dot{\rho}_0 \dot{z}_0}{\rho_0^2}
\right ) \nonumber \\
& &  + \nUo \left ( \dot{\X}_1 - 2 \X_1^2 + 2 \X_0^2 +
\frac{\dot{\rho}_0}{\rho_0} \X_1  - \frac{2 \dot{z}_0}{\rho_0} \X_0 +  
\frac{2 \dot{z}_0^2}{\rho_0^2} \right ). \label{MnnnUo}
\end{eqnarray}
\end{widetext}
Substituting the transverse-transverse derivatives (\ref{MnnUo}),
(\ref{MnnUprime}), and (\ref{MnnnUo}) into (\ref{Mnfper}),
(\ref{Mfperper}) and (\ref{MAllFo}) with the
substitutions $\F \rightarrow U$, $\H \rightarrow \G$ and $\J
\rightarrow \L$ we find the Cauchy $U$-boundary data in Proposition
\ref{FullBoundary}. Note that in this evaluation we need to use the formulae
for $\Uprimeo$ and $\nUprimeo$ in those for $\Uprimeprimeo$ and
$\nUprimeprimeo$.  Finally, the substitution $\F \rightarrow \Omega$,
$\H \rightarrow \Y$ and $\J \rightarrow \W$ and use of
(\ref{MnnOmegaprime}) and the fact that $\Omega_0 = 0$ gives us the
$\Omega$-boundary data. This completes the proof of the Proposition.


\end{document}